\documentclass[article]{JFM}
\usepackage{graphicx,natbib,tabularx,bm,amsmath}
\usepackage{subfigure}

\title[Multi-scale Grid Turbulence]{Freely-Decaying, Homogeneous Turbulence Generated by Multi-scale Grids}

\author[P.-\r{A}. Krogstad and P.A. Davidson]
{P.\ls-\ls \r{A}.\ls K\ls R\ls O\ls G\ls S\ls T\ls A\ls D$^1$
\and P.\ls A.\ls D\ls A\ls V\ls I\ls D\ls S\ls O\ls N$^2$}

\affiliation{$^1$Norwegian University of Science and Technology, N-7491 Trondheim, Norway.\\[\affilskip]
$^2$University of Cambridge, Cambridge CB2 1PZ, U.K.\\[\affilskip]
}

\date{29 March 2011}
\begin{document}
\maketitle

\begin{abstract}
We investigate wind tunnel turbulence generated by both conventional and
multi-scale grids. Measurements were made in a tunnel which has a
large test-section, so that possible side wall effects are very small and the length assures that the turbulence has time to settle down to a homogeneous shear-free state. The conventional and multi-scale grids
were all designed to produce turbulence with the same integral scale, so
that a direct comparison could be made between the different flows. Our
primary finding is that the behavior of the turbulence behind our
multi-scale grids is virtually identical to that behind the equivalent
conventional grid. In particular, all flows exhibit a power-law decay of
energy, $u^2 \sim t^{-n}$, where $n$ is very close to the classical Saffman exponent
of $n = 6/5$. Moreover, all spectra exhibit classical Kolmogorov scaling,
with the spectra collapsing on the integral scales at small $k$, and on the
Kolmogorov micro-scales at large $k$. Our results are at odds with some other
experiments performed on similar multi-scale grids, where significantly higher energy decay exponents and turbulence levels have been reported.
\end{abstract}

\section{\label{intro}Introduction}
In recent years a number of wind-tunnel experiments and numerical simulations have focused on quasi-homogeneous turbulence generated by multi-scale grids. (See, for example, \citeauthor{Hurst}, \citeyear{Hurst}, and \citeauthor{Nagata}, \citeyear{Nagata}.) When the grids concerned possess certain fractal-like properties, this is sometimes referred to as fractal-generated turbulence.  In some of these experiments the turbulence appears to behave in unexpected ways. For example, \cite{Hurst} report unusually high values of  $Re_{\lambda} = u \lambda / \nu$, perhaps twice that associated with the equivalent conventional grid. (Here $\lambda$ is the Taylor micro scale, $\lambda^2 = \left< u_x^2 \right> / \left< \left( \partial u_x / \partial x \right)^2 \right>$, $\nu$ the viscosity, and $u = \sqrt{ \left< {\mathbf u}^2 \right> / 3}$ the characteristic fluctuating velocity.) The same paper also reports energy decay characteristics, such as power-law decay, $ \left< {\mathbf u}^2 \right> \sim t^{-n}$ , with very high decay exponents ($n \sim 2$ rather than the classical $n \sim 1.2 \rightarrow 1.4$), and also exponential decay rate for some grid geometries. This is surprising as there are good arguments why classical fully-developed homogeneous turbulence can decay no faster than $n \sim 10/7$ (see e.g.  Appendix \ref{appendix} of this paper, \citeauthor{Krogstad}, \citeyear{Krogstad}, or \citeauthor{Ossai}, \citeyear{Ossai}, for a summary of these arguments). The implication is that turbulence generated by multi-scale excitation can exhibit non-classical characteristics, which has led to a re-examination of the assumtions underlying energy decay laws (see for example \citeauthor{George}, \citeyear{George}, and \citeauthor{Hosokawa}, \citeyear{Hosokawa}). 

However, this suggestion is somewhat at odds with the evidence of direct numerical simulations in periodic cubes, where it is usually found that, after some transient, the behavior of the turbulence is largely independent of the precise form of the initial energy spectrum (see, for example,  \citeauthor{Ossai}, \citeyear{Ossai}). Indeed, it is often argued that, once fully developed, all the turbulence remembers of its initial conditions is the prefactor, $c_m$, in the expression $E(k \rightarrow 0) = c_mk^m$, $c_m$ being an invariant for $m \leq 4$  (\citeauthor{Ossai}, \citeyear{Ossai}, \citeauthor{davidson04}, \citeyear{davidson04}, \citeauthor{Ishida}, \citeyear{Ishida}, and \citeauthor{davidson09}, \citeyear{davidson09}).

There is, however, a different interpretation of the experiments of \cite{Hurst}, one which is consistent with the evidence of the numerical simulations. Immediately behind a multi-scale grid the turbulence is far from homogeneous, and there is a significant production of turbulent energy due to transverse gradients in the mean velocity. Moreover, this inhomogeneity in the mean and fluctuating velocities is stronger than for a conventional grid and it persists for many tens of integral scales downstream of the mesh (\citeauthor{Nagata}, \citeyear{Nagata}). This might not matter if all measurements are made well downstream of a multi-scale grid, where approximate homogeneity is recovered. In conventional grid turbulence it is considered good practice to restrict measurements to, say, $x > 25 \rightarrow 30M$ in order to avoid the inhomogeneous region behind the grid (e.g. \citeauthor{Comte-Bellot}, \citeyear{Comte-Bellot}, \citeauthor{Gad}, \citeyear{Gad}). It is possible, therefore, that the apparently non-classical results obtained for some of the grid geometries of \cite{Hurst} are a consequence of a distinct lack of homogeneity immediately behind the grid, i.e. a transient phenomenon which disappears further downstream. Indeed, \citeauthor{Hurst}, \citeyear{Hurst}, themselves note that more extensive tests are required in order to assess the role of inhomogeneities close to the grid. 

In order to test the hypothesis that turbulence behind multiscale grids will develop in a classical way (after an initial transient), we have conducted a series of wind tunnel experiments using two multi-scale grids of the cross class described in \cite{Hurst}, and compared these results with data obtained using a conventional grid. In order to ensure that the comparison is meaningful, the dimensions of the grids were chosen such that the integral scale of the turbulence some short distance downstream of each grid, $\ell_0$, is virtually the same in all three cases (to within a percent or so). Crucially, the experiments were carried out in a large working section, and measurements taken over the extended range $80 \ell_0 < x < 400 \ell_0$, which translates to $50 M < x < 240M$  for the conventional grid. Excellent homogeneity was obtained throughout the measurement range for all three grids and the long working section allowed us to determine the energy decay characteristics with considerable accuracy. As we shall see, the fully-developed turbulence generated by all three grids is virtually identical. In particular, there is no significant difference in the behavior of  $Re_{\lambda}$, and in all cases the decay exponent $n$ in the expression $\left< {\mathbf u}^2 \right> \sim t^{-n}$  is very close to the classical prediction of \cite{Saffman}, i.e. $n = 6/5$. (For a discussion of decay exponents in classical turbulence see e.g. \citeauthor{davidson04}, \citeyear{davidson04} and \citeauthor{Krogstad}, \citeyear{Krogstad}.)

\section{\label{Exp}The Experimental Set-up}

The experiments were performed in the large recirculating wind tunnel described in \cite{Krogstad}. The tunnel test-section has transverse dimensions of 2.7m x 1.8m (measured at the start of the test section) and is 12m long. There is an adjustable roof to compensate for the growth of the sidewall boundary layers and the grids were mounted upstream in the test section contraction to improve isotropy. From the location of the grid to the entrance of the test section, the area contraction ratio was 1.48 and the test section starts $x = 1.2m$ downstream of the grid.

All three grids were produced from 2mm thick sheet metal. The conventional grid (labeled $cg$) has square holes 30mm x 30mm punched at 40mm spacing, giving a mesh size of $M = 40mm$, a bar width of $t = 10mm$, and a solidity of $\sigma = 44\%$. The tests on this grid were all performed at a Reynolds number of $Re_M = U M / \nu = 3.6 \times 10^4$, where $U$ = 13.5m/s was the mean speed in the tunnel. 

\begin{figure}
\begin{center}
  \hspace{-3mm}
\subfigure{
  \includegraphics[width=.5\textwidth]{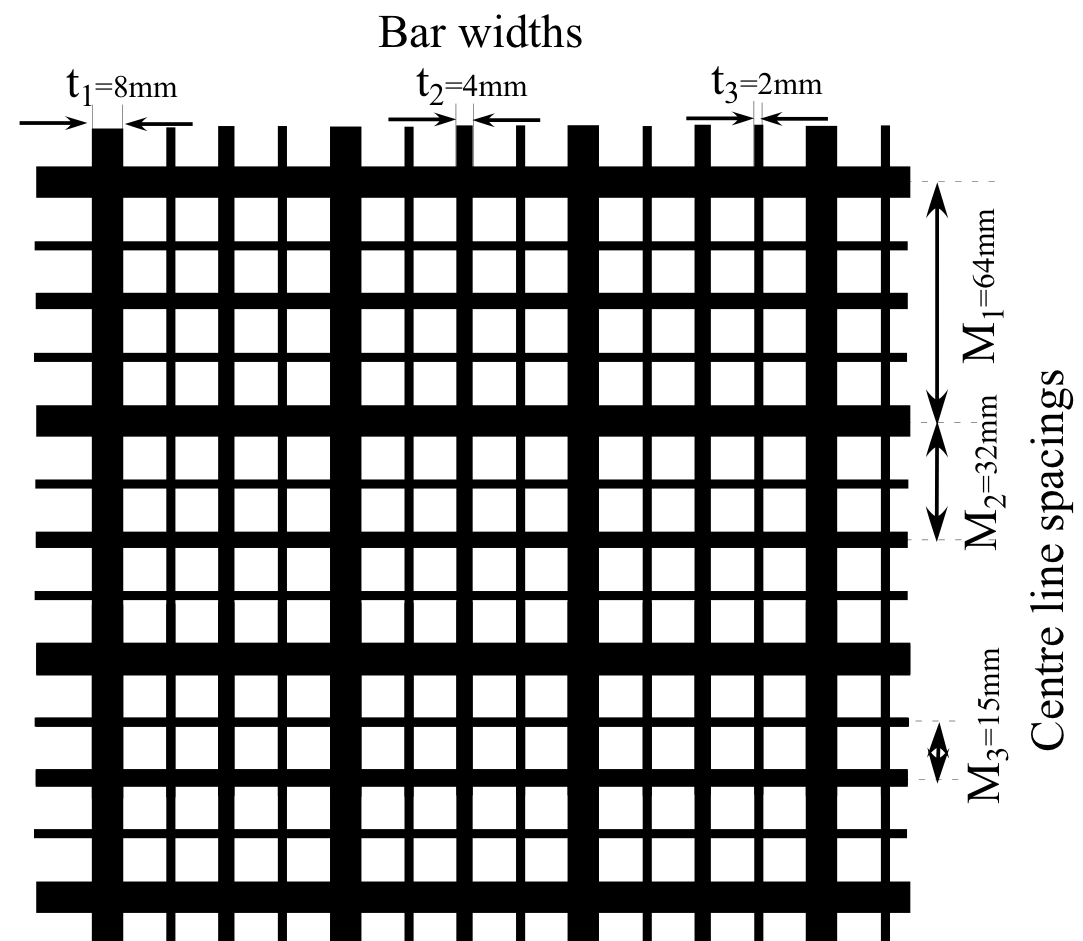}
  \label{fig:Fig1a}}
\hspace{2mm}
\subfigure{
  \includegraphics[width=.45\textwidth]{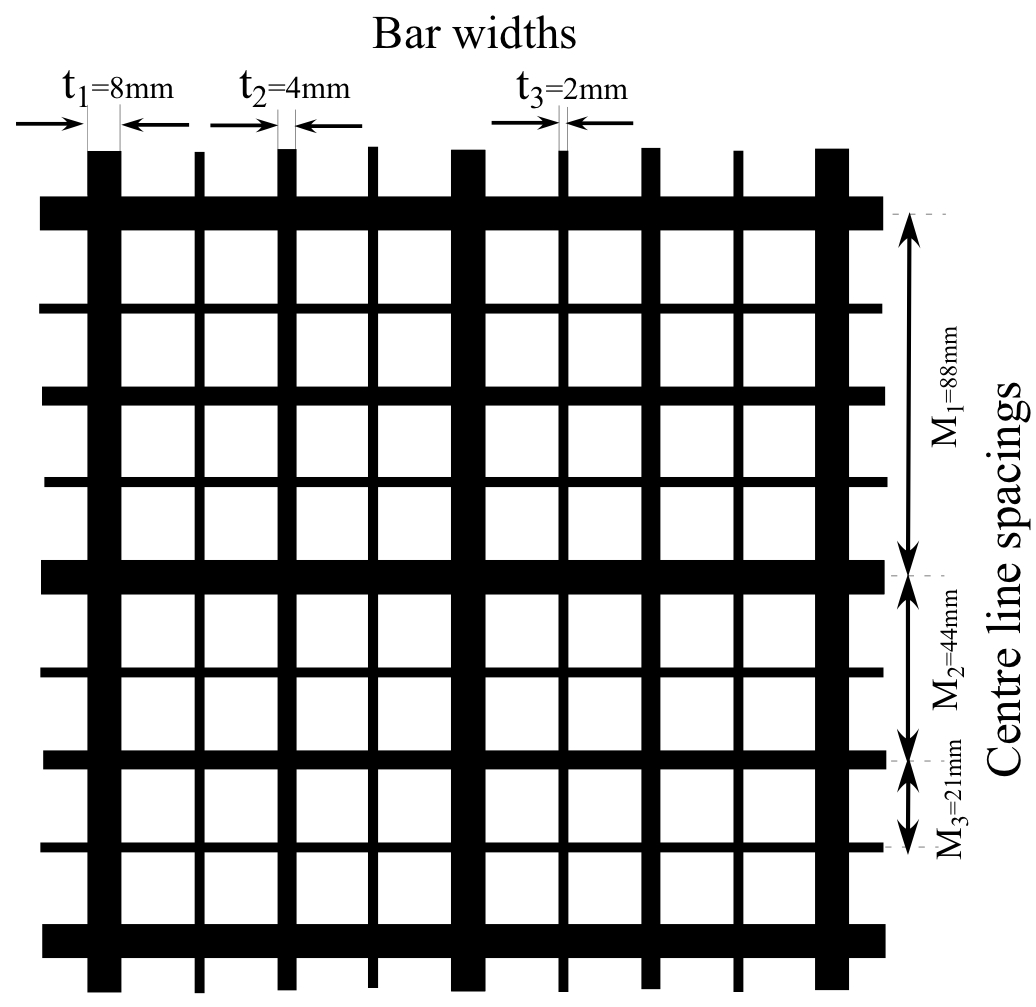}
  \label{fig:Fig1b}}
  \vspace{-2mm}
  \hspace{-4mm}
(a)
  \hspace{61mm}
(b)
  \caption{The two multi-scale grids used in this study.}
 \label{fig:Grids}
 \end{center}
\end{figure}

The first of the multi-scale grids (labeled $msg1$) is similar to the cross-grid-type($a$) of  \cite{Hurst} (which we shall label $cg-a$) and is shown in Figure \ref{fig:Fig1a}. It has bar widths ranging from $t_1 = 8mm$ down to $t_3 = 2mm$, and mesh sizes ranging from $M_1 = 64mm$ to $M_3 = 15mm$. The solidity of $msg1$ is also $\sigma = 44\%$. These measurements were taken at $U$ = 14.0m/s. 

The second multi-scale grid ($msg2$)  is shown in Figure \ref{fig:Fig1b}. As for $msg1$, the bar widths vary from $t_1 = 8mm$ down to $t_3 = 2mm$, though the mesh sizes are larger, with $M_1 = 88mm$ to $M_3 = 21mm$. This reduces the solidity of $msg2$ to $\sigma = 33\%$. This grid was tested at $U$ = 15.5m/s.

As we shall see, the turbulence produced by these grids becomes more-or-less homogeneous and fully developed at around $x = 2m$, at which point the Kolmogorov microscale is $\eta \approx 0.22mm \rightarrow 0.26mm$. On the other hand, the integral scales at $x = 2m$, $\ell_0 = \ell(x=2m)$, obtained in the usual way by integrating the longitudinal  correlation function
\begin{equation}
 \ell = \int_0^{\infty} \frac{\left< u_x (x) u_x (x+r)\right>}{\left< u_x^2\right>}dr \; ,
\label{Eq2-1}
\end{equation}
turn out to be $\ell_0 = 23.9mm$ (for $cg$), $\ell_0 = 23.6mm$ (for $msg1$), and $\ell_0 = 23.4mm$ (for $msg2$), respectively. (The measurements of $\eta$ and $\ell$ are discussed in $\S$\ref{Development}.) Note the uniformity of $\ell_0$ across the grids. Note also that the geometric length-scales associated with the two multi-scale grids almost span the range of dynamic scales associated with the turbulence, from around $9\eta$ up to several integral scales. Finally we note that, in terms of $\ell_0$, the tunnel cross-section is approximately $115\ell_0 \times 80\ell_0$, thus ensuring that there is minimal influence of the side-wall boundary layers. Measurements are taken up to a considerable distance downstream of the grid, allowing an accurate estimate of the energy decay exponent, $n$.  

The data was obtained using single and two component hot-wire anemometry. For the measurements of the decay of $\left< \left( u_x \right)^2 \right>$, spectra, the length scales $\eta$ and $\ell$ etc. a purpose made $2.5 \mu m$ partly etched Platinum-10\%Rhodium straight single wire probe was produced. The active wire length was $0.5mm$. For two component measurements and for global checks such as spanwise homogeneity etc. an X-wire probe with $\pm 45^o$ nominal wire angles was produced. For this probe $5 \mu m$ partly etched wires were used with $1mm$ wire lengths. The two wires were also separated in the spanwise direction by $\Delta z = 1mm$. By careful handling of the two delicate probes it was possible to maintain the same wires throughout the whole experiment. Hence there should be no effects in the data that can be attributed to changes in probe characteristics.

The probes were operated at an overheat temperature of about 320 degrees using in-house manufactured anemometers which were tuned to a frequency response $f_w$ of at least 20kHz. ($f_w$ is here calculated from $f_w = (1.3 \tau)^{-1}$, where $\tau$ is the time for the anemometer output to recover to within 10\% of the asymptotic level after a square wave step change has been inserted at the top of the Whetstone bridge in the anemometer.) The output from the anemometers were suitably offset and amplified to span as much as possible of the $\pm$10 volt range of the 16 bit acquisition card used before the signals were passed through an AC coupled Krohn-Hite amplifier and low-pass filter unit. The filter frequency $f_c$ was set by inspecting the dissipation spectra of a few initial measurements with very high filter settings. $f_c$ was then set at the frequency where noise first started to affect the dissipation spectra and a new set of data was obtained at a sampling frequency which was slightly higher than 2$f_c$. 

For verification purpose only, a number of two component laser Doppler anemometry (LDA) measurements were taken for each grid, but these data were not used in the data analysis presented here, since LDA tends to produce too high turbulence levels as the turbulence intensity drops below about 1\%. Further information about the instrumentation and data analysis may be found in \cite{Krogstad}.

\section{\label{Isotropy}Tests for Homogeneity and Isotropy.}

\begin{figure}
\begin{center}
  \hspace{-3mm}
  \includegraphics[width=1.05\textwidth]{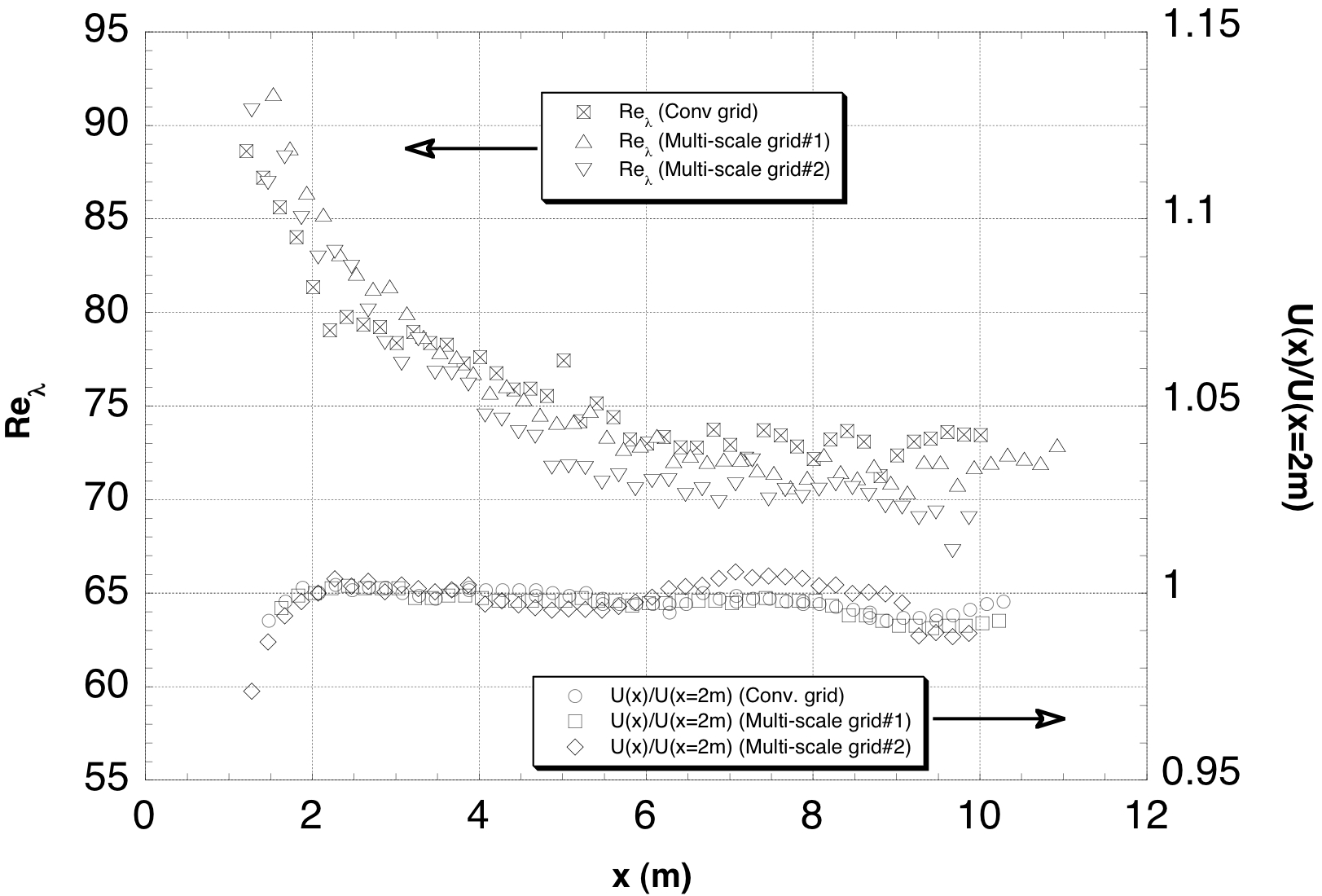}
  \label{fig:Fig2}
  \caption{Streamwise distributions of $Re_{\lambda}$ (left axis) and the test section centre-line speed, $U$, (right axis). $U$ is normalized by $U$ at $x=2m$.}
 \label{fig:Re_lambda}
 \end{center}
\end{figure}

Let us start by documenting the relative degree of homogeneity and isotropy in the experiment. Figure \ref{fig:Re_lambda} shows the time-averaged centre-line velocity, normalized by $U_0 = U(x=2m)$, for all three grids as a function of $x$. For $x> 2m$,  $U$ is constant to within $\pm 0.5\%$, with $Re = U \ell_0 / \nu = 2.1 \times 10^4$, $2.2 \times 10^4$ and $2.4 \times 10^4$, respectively, for the three grids. However, for $x < 2m$  there is a noticeable streamwise acceleration due to the effect of the contraction extending slightly into the first part of the test section.  In the light of the streamwise acceleration, we shall ignore data for $x < 2m$  (i.e. $x < 80\ell_0$). Figure \ref{fig:Re_lambda}  also shows the streamwise variation of $Re_{\lambda} = \left< u_x^2 \right>^{1/2} \lambda / \nu$ for all three grids. The present values are only about one third of those reported by \cite{Hurst} and there is no significant difference between the conventional and multi-scale grids.

\begin{figure}
\begin{center}
\subfigure{
(a)
  \includegraphics[width=.8\textwidth]{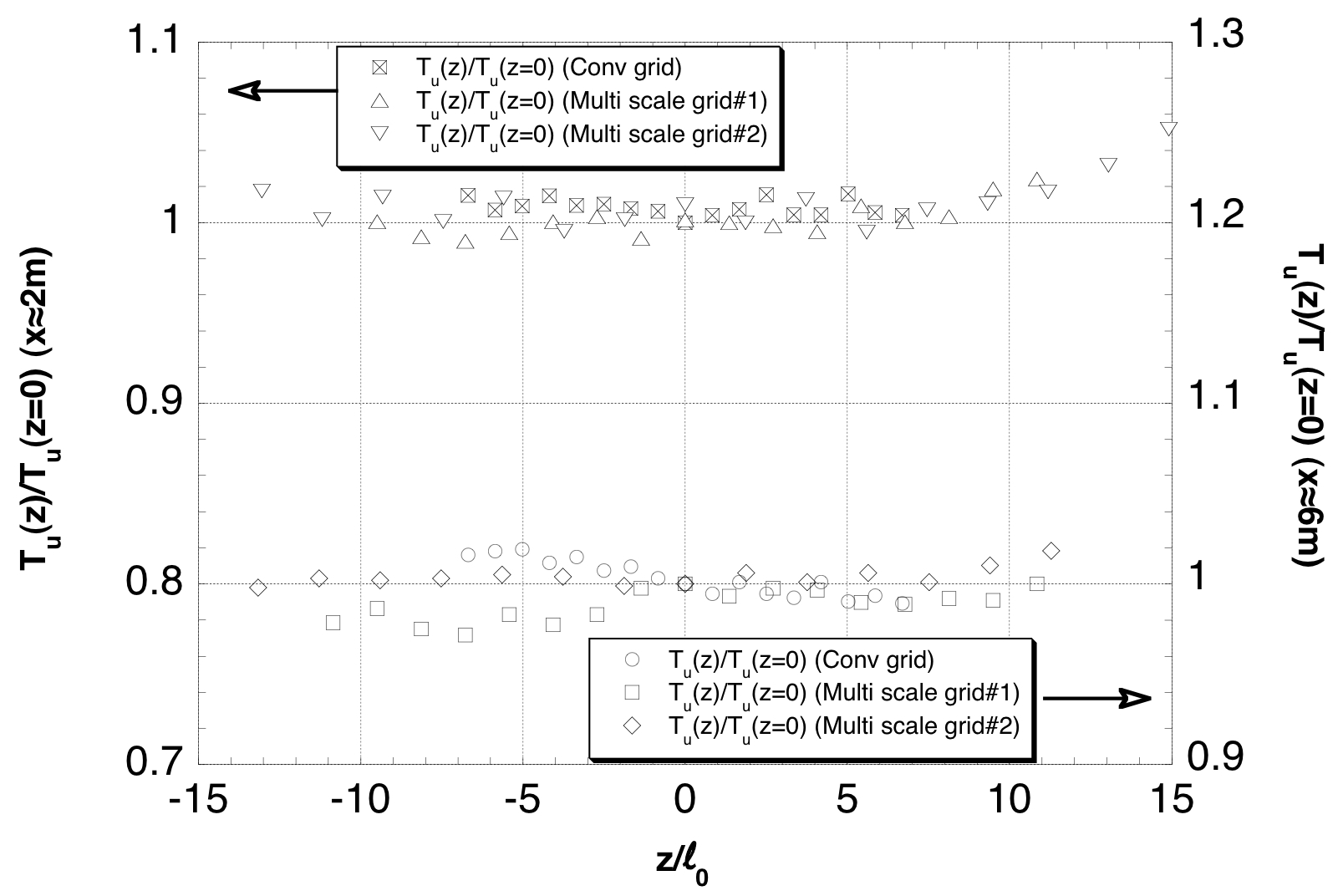}
  \label{fig:Fig3a}}\\
\subfigure{
(b)
  \includegraphics[width=.8\textwidth]{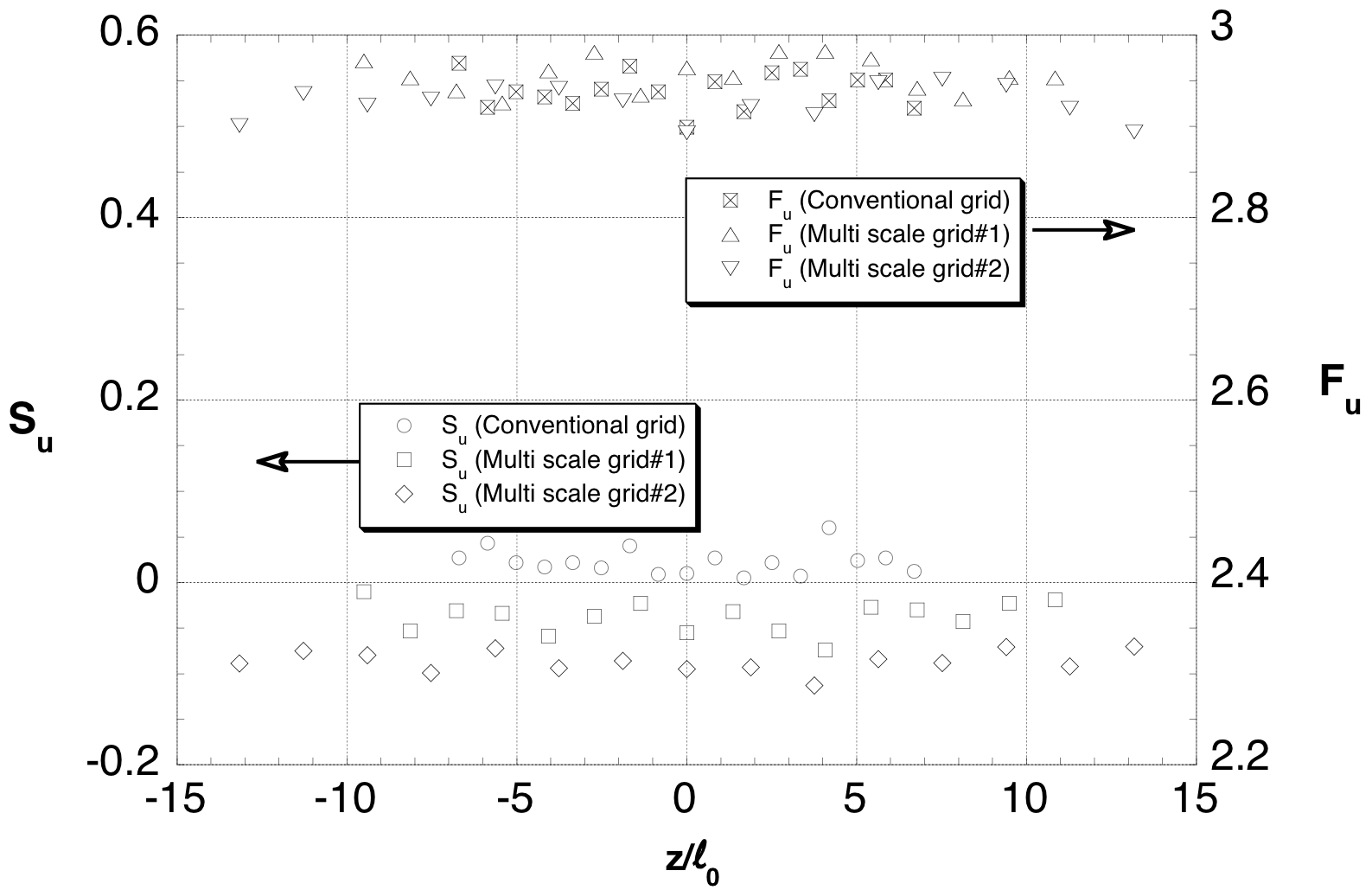}
  \label{fig:Fig3b}}\\
  \caption{Spanwise distributions of: (a) $T_u$ at $x \approx 2m$ (left axis) and $x \approx 6m$ (right axis), and (b) skewness, $S_u$, of $u_x$ at   (left axis) and flatness, $F_u$, of $u_x$ at $x = 2m$ (right axis).}
 \label{fig:Spanwise}
 \end{center}
\end{figure}

The spanwise distributions of the turbulence intensity, $T_u = u/U$, where $u = \left< u_x^2 \right>^{1/2}$, measured at $x \approx 2m$ and $x \approx 6m$ are shown in Figure \ref{fig:Spanwise}, along with the spanwise distribution of the skewness, $S_u$, and flatness, $F_u$, of $u_x$ at $x \approx 2m$. The measurements were made over a span corresponding to about $z \approx \pm 4M$, where $M$ represents the largest mesh of each grid. Since $M$ is only a well defined scaling length for the conventional grid, we have chosen to scale $z$ with $\ell_0$, hence the apparent differences in the measurement ranges in the figure.

Again, all three grids are seen to behave in a similar manner, with variations in $T_u $ limited to about $\pm 3\%$ at $x = 2m$ and $\pm 2\%$ at $x = 6m$. The skewness and  flatness distributions are also well behaved in all three cases, with $S_u$ being slightly negative and $F_u \approx 2.95$, values which are typical of most grid turbulence experiments. We conclude that, for $x > 2m$, the turbulence is relatively homogeneous. 

\begin{figure}
\begin{center}
  \hspace{-3mm}
  \includegraphics[width=1.0\textwidth]{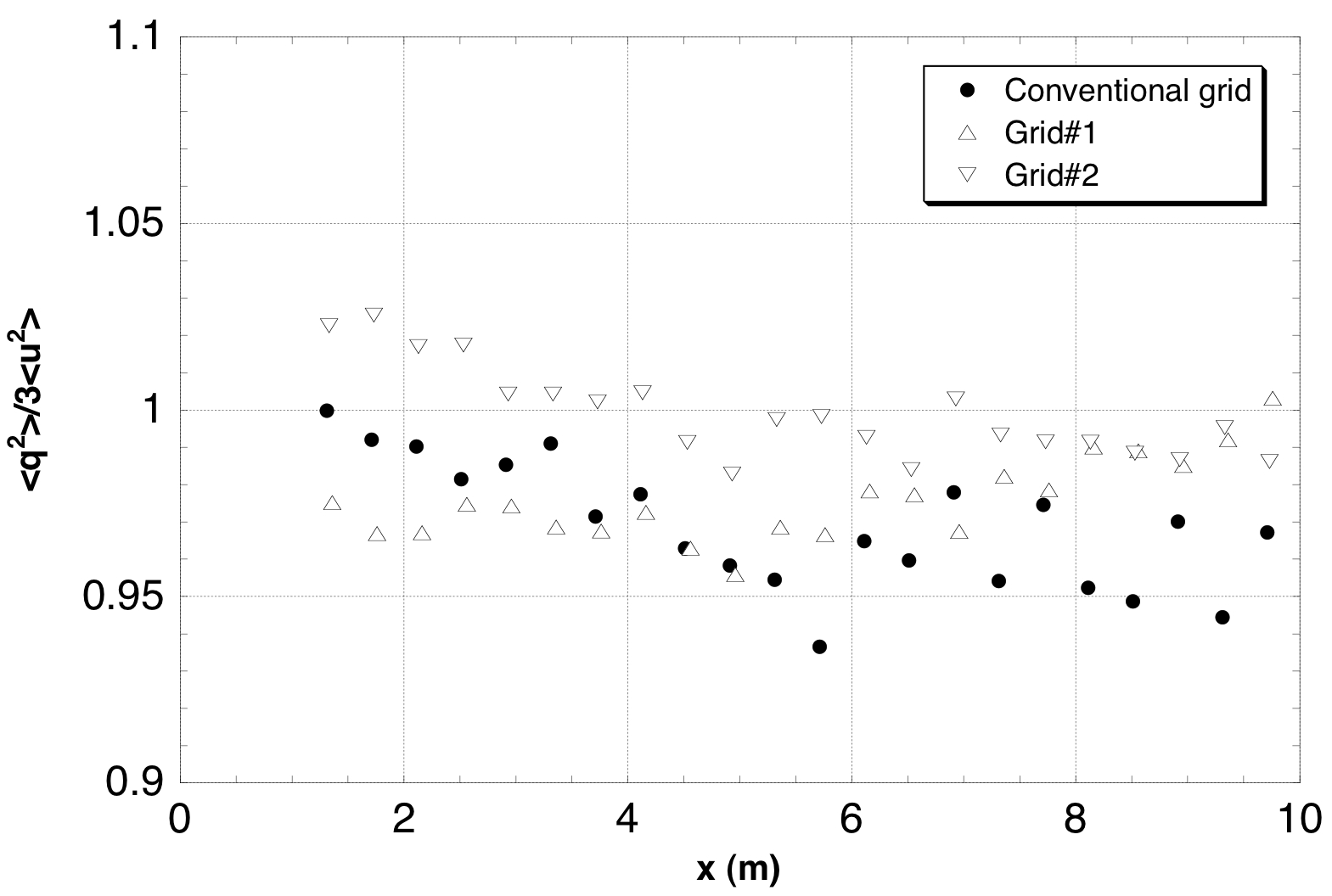}
  \caption{Streamwise distributions of  $\left< q^2 \right> / 3\left< u_x^2 \right>$.}
 \label{fig:q2}
 \end{center}
\end{figure}

The streamwise development of the anisotropy of the turbulence, as measured by the ratios $\left< u_x^2 \right> / \left< u_y^2 \right>$, $\left< u_x^2 \right> / \left< u_z^2 \right>$ and $\left< q^2 \right> / 3\left< u_x^2 \right>$, where $\left< q^2 \right> = \left< {\mathbf u}^2 \right> = \left< u_x^2 + u_y^2 + u_z^2 \right> $, was measured for each grid (Figure \ref{fig:q2}). For all cases it was found to follow closely the distributions reported for the conventional grid in Figure 3 of \cite{Krogstad}. We show here just the three streamwise developments of  $\left< q^2 \right> / 3\left< u_x^2 \right>$ which is seen to be very similar for the 3 cases and the ratio is very close to the isotropic value of 1. For all grids $\left< u_x^2 \right> / \left< u_y^2 \right>$, $\left< u_x^2 \right> / \left< u_z^2 \right>$ and $\left< q^2 \right> / 3\left< u_x^2 \right>$  were close to unity at $x = 2m$, but $\left< u_x^2 \right> / \left< u_y^2 \right>$ and $\left< u_x^2 \right> / \left< u_z^2 \right>$ were found to divert slowly with increasing $x$. The largest departure from isotropy was observed at the exit of the test section, where $\left< u_x^2 \right> / \left< u_z^2 \right>$ reached values in the range of 0.8 to 0.9. There was a corresponding growth in $\left< u_x^2 \right> / \left< u_y^2 \right>$, to produce the almost constant $\left< q^2 \right> / 3\left< u_x^2 \right>$ ratios shown in the figure. 

\begin{figure}
\begin{center}
  \hspace{-3mm}
  \includegraphics[width=1.0\textwidth]{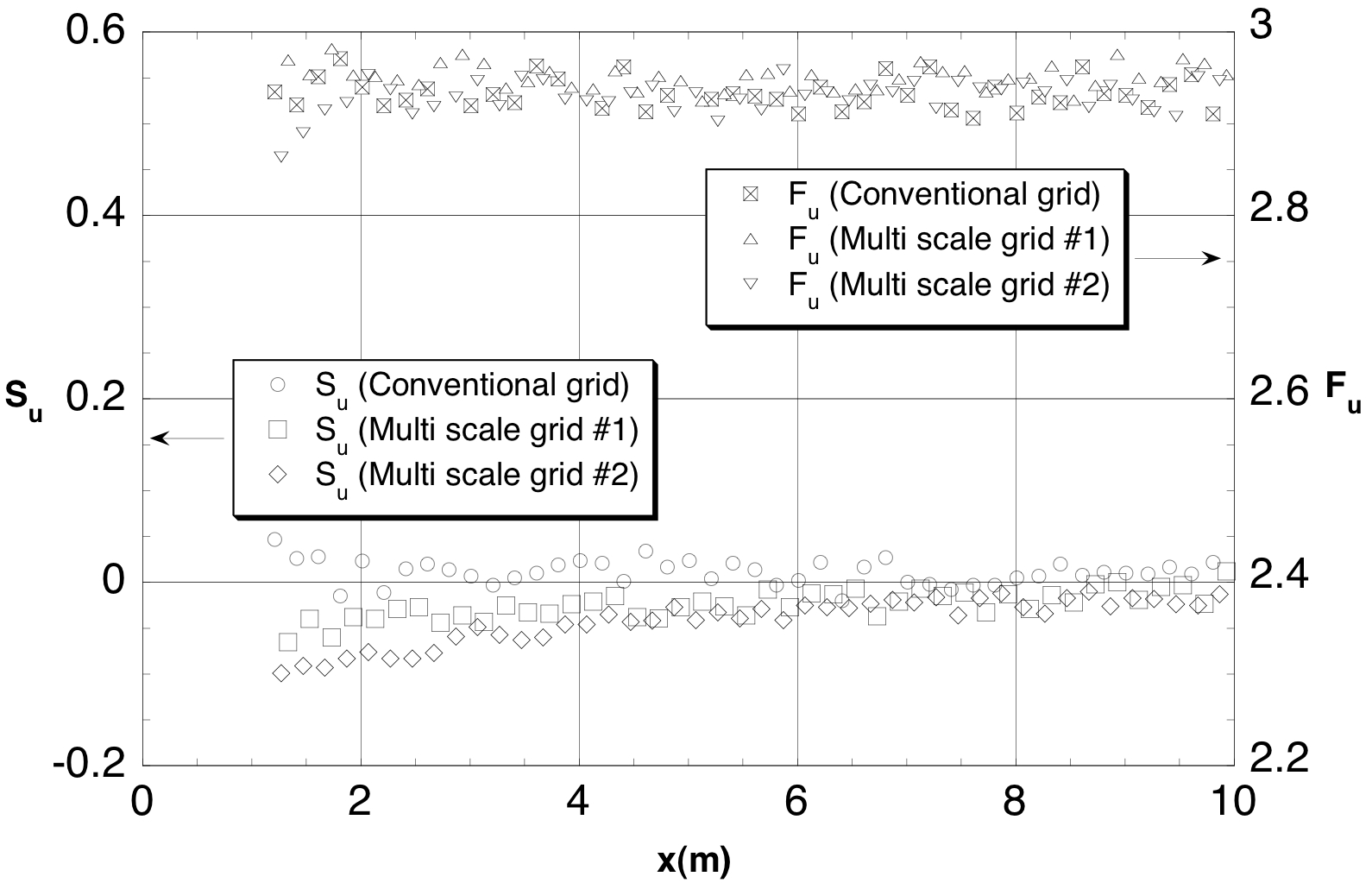}
  \caption{Streamwise distributions of  $\left< q^2 \right> / 3\left< u_x^2 \right>$.}
 \label{fig:SuFu}
 \end{center}
\end{figure}

Figure \ref{fig:SuFu} shows the skewness and flatness distributions of the streamwise fluctuations measured along the centre line of the test section. The flatness is constant throughout the measurement range at $F_u \approx 2.95$ for all grids. The skewness is slightly different for the grids initially, as was shown in Figure \ref{fig:Fig3b}, but they all tend steadily towards $S_u \rightarrow 0$ as the flow develops downstream.

In most of what follows we shall restrict the analysis of our data to the region $x > 2m$, to avoid any acceleration effects caused by the inhomogeneities in $U$, and to $x < 8m$, because of the increased levels of noise for large $x$ as the turbulence level drops well below 1\%. For completeness, however, the full set of data will be included in the plots so that the reader may see where and how the data departs from classical decay behavior. From the data presented above we see that, in this range, the turbulence is reasonably homogeneous and that, although there is some anisotropy, the levels are not excessive and are comparable to, if not better, than in most other experiments.

\section{\label{Decay}The Energy Decay Rate.}

We shall now estimate the decay exponent, $n$, in the decay law $\left< {\mathbf u}^2 \right> \sim t^{-n}$ for all three grids. First, however, perhaps it is worth saying something about the values this exponent might be expected to take on theoretical grounds. As noted in \citeauthor{davidson04} (\citeyear{davidson04}, \citeyear{davidson09}) and \cite{Krogstad}, there are two classical predictions for $n$. One arises when $E(k \rightarrow 0) \sim k^2$, a situation called Saffman turbulence, and the other when $E(k \rightarrow 0) \sim k^4$, so-called Batchelor turbulence. In the former case it may be shown that
\begin{equation}
L = \int \left< {\mathbf u} \cdot {\mathbf u'} \right> d{\mathbf r} = {\rm constant} \; ,\; \; \; ({\rm for} \;  E(k) \sim k^2)  \; ,
\label{Eq4-1}
\end{equation}
where $\left< {\mathbf u} \cdot {\mathbf u'} \right>$ is the usual two point velocity correlation (see \citeauthor{Saffman}, \citeyear{Saffman}), while in the latter case
\begin{equation}
I = - \int r^2 \left< {\mathbf u} \cdot {\mathbf u'} \right> d{\mathbf r} = {\rm constant \; ,\; \; \;  (for} \;  E(k) \sim k^4)  \; ,
\label{Eq4-2}
\end{equation}
with the caveat that Eq. (\ref{Eq4-2}) holds only once the turbulence is fully-developed (see \citeauthor{Ishida}, \citeyear{Ishida}). These integrals are dominated by the large scales and so self similarity of the large scales (when it applies) demands 
\begin{equation}
u^2 \ell^3 = {\rm constant \; ,\; \; \;  (for \;  Saffman \;  turbulence)}  \; ,
\label{Eq4-3}
\end{equation}
\begin{equation}
u^2 \ell^5 = {\rm constant \; ,\; \; \;  (for \;  Batchelor \;  turbulence)}  \; .
\label{Eq4-4}
\end{equation}

These may be combined with the empirical, but well supported, law
\begin{equation}
\frac{d u^2}{dt} = - A \frac{u^3}{\ell} \;,  \; \;  A = {\rm constant }  \; ,
\label{Eq4-5}
\end{equation}
to give $n = 6 / 5$  (Saffman's exponent; see \citeauthor{Saffman}, \citeyear{Saffman}) in $E(k) \sim k^2$ turbulence, and $n = 10 / 7$ (Kolmogorov's exponent; see \citeauthor{Kolmogorov}, \citeyear{Kolmogorov}) in $E(k) \sim k^4$ turbulence. The first of these exponents was observed in the experiments of \cite{Krogstad} and the second in the numerical simulations of \cite{Ishida}. While other decay exponents have been proposed from time to time, it is natural to keep these two classical predictions in mind when examining the experimental data. However, there is a slight complication which arises when comparing these predictions with experiments: in wind tunnel data the coefficient $A$ can vary slowly along the test section, and this causes slight departures from the ideal values of $n = 6 / 5$ or $n = 10 / 7$, even when Eq. (\ref{Eq4-3}) or (\ref{Eq4-4}) hold true (see \citeauthor{Krogstad}, \citeyear{Krogstad}). Indeed, we shall see shortly that just such a slow variation of $A$ occurs in our experiments. 

Let us now estimate the decay exponents for our three grids by comparing the experimental data with the power law
\begin{equation}
\frac{\left< u_x^2 \right>}{U^2} = a \left[ \frac{x-x_0}{\ell_0}Ê\right]^{-n}  \; ,
\label{Eq4-6}
\end{equation}
where, as before, $\ell_0$ is the integral scale at $x = 2m$. Perhaps we should note from the outset that it is notoriously difficult to obtain reliable estimates of $n$. There are several reasons for this. First, as shown in \cite{Krogstad}, the virtual origin $x_0$ in Eq. (\ref{Eq4-6}) does not correspond to the point where the turbulence first matures, but rather is located somewhat upstream of that position, and we do not know in advance where $x_0$ will lie. Second, if the range of $x / \ell_0$ is too short, the decay exponent becomes very sensitive to the choice of the unknown $x_0$. Third, if data from the inhomogeneous region close to the grid is included in the fit, then higher (and misleading) values of $n$ are usually obtained. Fourth, for large $x / \ell_0$ the turbulence intensity is low, so that noise starts to become problematic, obvious sources being electronic noise and pressure fluctuations arising from the side-wall boundary layers. In summary, then, to obtain reliable estimates of $n$ we require that: (i) $x_0$ be chosen systematically and with care; (ii) there is an extended $x / \ell_0$ range; (iii) anomalous data close to the grid must be excluded; and (iv) data for large $x / \ell_0$ should be ignored when the noise level becomes excessive. 

\begin{figure}
\begin{center}
  \hspace{-3mm}
  \includegraphics[width=1.0\textwidth]{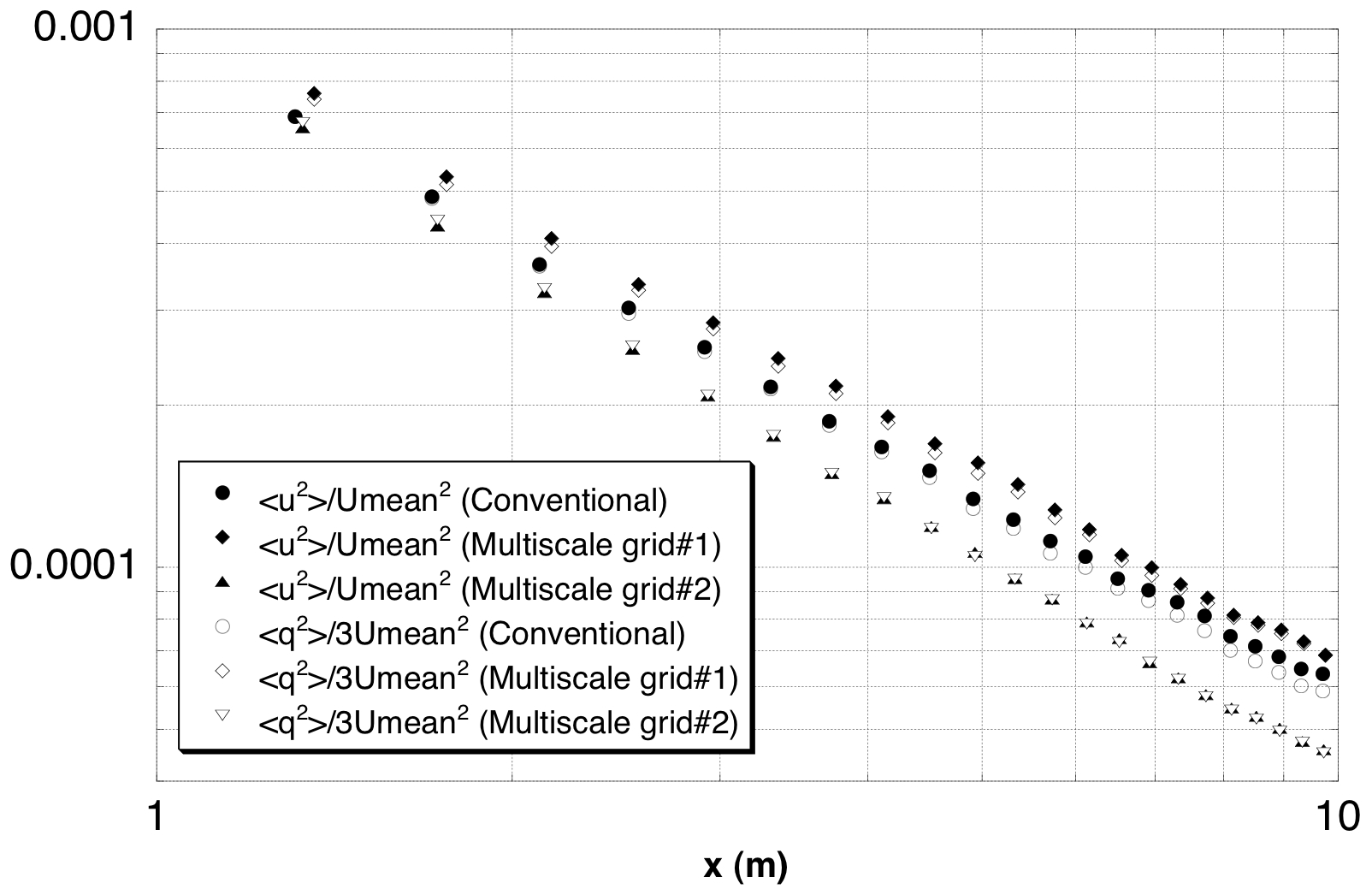}
  \label{fig:Fig5}
  \caption{Streamwise distributions of  $\left< u_x^2 \right>/U^2$ (filled symbols) and $\left< q^2 \right>/3U^2$ (open symbols).}
 \label{fig:decay}
 \end{center}
\end{figure}

Figure \ref{fig:decay} shows $\left< u_x^2 \right>/U^2$ and $\left< q^2 \right>/3U^2$, obtained using two different alignments of the x-wires, all plotted as a function of $x$. This log-log plot demonstrates that, for all three grids, there is a clear power-law relationship between $u^2$ and $x$. It also shows that $\left< u_x^2 \right>$ and $\left< q^2 \right>$  follow essentially the same power law. Instead of attempting a direct power-law fit to this data, it is convenient to rewrite Eq. (\ref{Eq4-6}) as 
\begin{equation}
{\rm ln} \left[ \frac{\left< u_x^2 \right>}{U^2} \right] = {\rm ln} a - n {\rm ln} \left[ \frac{x-x_0}{\ell_0}Ê\right]  \; ,
\label{Eq4-7}
\end{equation}
to which a linear fit may be made. In view of the difficulty in obtaining reliable estimates of $n$, three different fitting procedures will be used and their results compared. All three methods are described in detail in \cite{Krogstad}.

\subsection{\label{Regression}Regression method}

Here we search for the fit that gives the smallest variance, $\sigma^2$, between the data and Eq. (\ref{Eq4-7}). The procedure works as follows. Starting with all of the data, a linear fit to Eq. (\ref{Eq4-7}) is made and $\sigma^2$ noted. Data at the most downstream location is then removed one by one until a minimum in $\sigma^2$ is obtained. This then determines $x_{max}$, the point beyond which noise starts to have a significant effect on the data. With $x_{max}$ fixed, a similar procedure is then applied at the most upstream location and used to determine $x_{min}$, the location of the start of the admissible data. The best fit values of $a$, $x_0$ and $n$ for these values of $x_{min}$ and $x_{max}$ are then taken to be the optimum values for these parameters. The exact locations of $x_{min}$ and $x_{max}$ vary slightly from grid to grid and between fitting to $\left< u_x^2 \right>$ and $\left< q^2 \right>$, but typically $x_{min} \approx 1.5m$ (i.e. $\sim 60\ell_0$) and $x_{max} \approx 8m$ (i.e. $\sim 330\ell_0$). 

\begin{figure}
\begin{center}
  \hspace{-3mm}
  \includegraphics[width=1.0\textwidth]{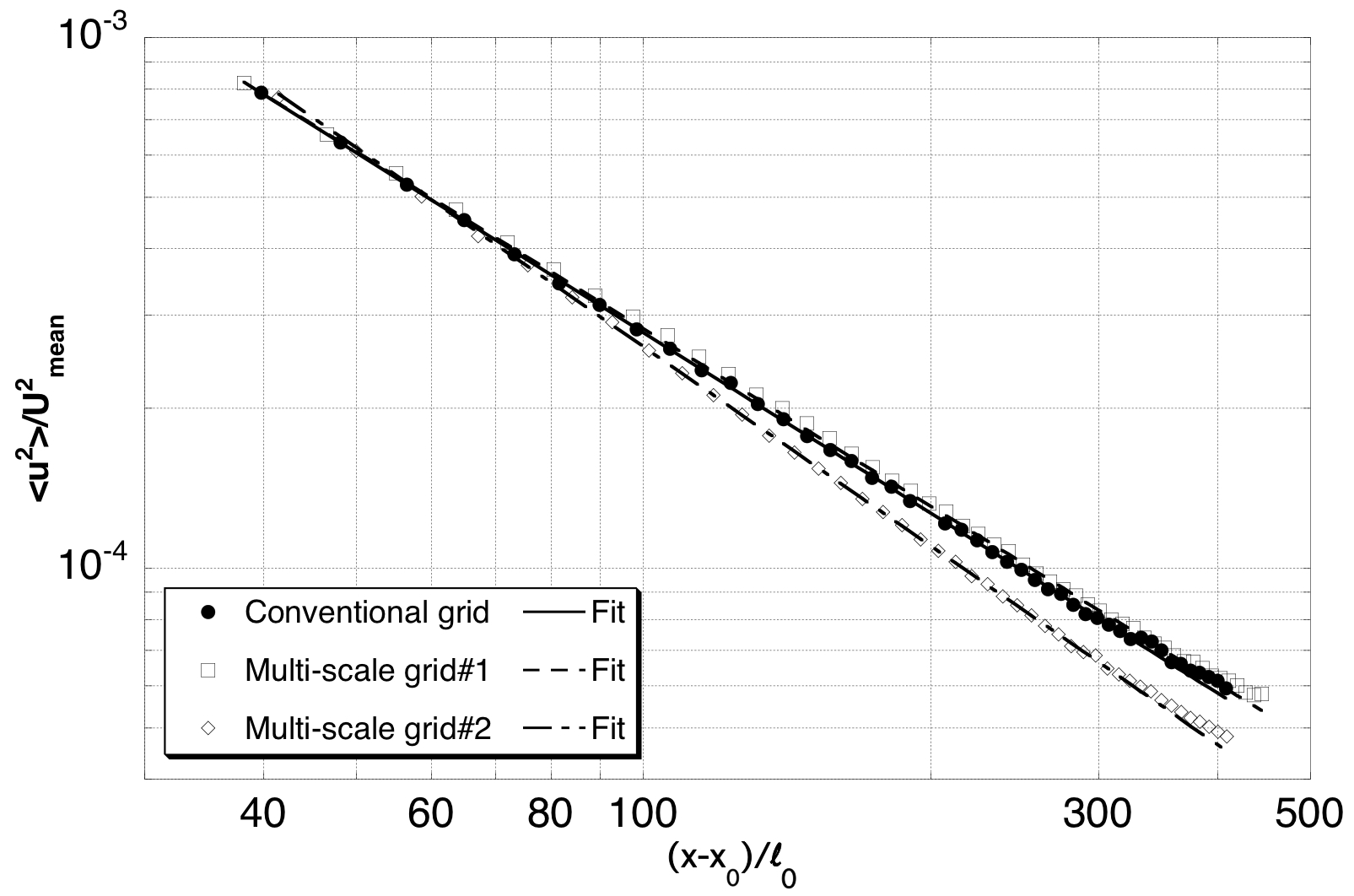}
  \label{fig:Fig6}
  \caption{$\left< u_x^2 \right>/U^2$ versus $\left( x-x_0 \right) / \ell_0$ and the corresponding best-fit power laws obtained using the regression method.}
 \label{fig:fit}
 \end{center}
\end{figure}

This procedure applied to $\left< u_x^2 \right>$ yields $n$ = 1.13, 1.11, and 1.25 for grids $cg$, $msg1$ and $msg2$ respectively, as shown in Figure \ref{fig:fit}. The corresponding virtual origins are at $x_0$ = 0.26m, 0.43m and 0.30m. Marginally different exponents were found when applying this method to $\left< q^2 \right>$, and the range of values obtained using regression are noted in Table 1, alongside the range of exponents estimated using the two procedures described in the following sections. 

\begin{table}
\begin{center}
\begin{tabular}{|c|c|c|c|c|}
\hline
 & $x_0$  & $n$ & $n$ & $n$ \\
 & regression & regression & local exponent & maximum \\
 & method  & method & method & decay range \\
\hline
Conventional grid & 0.26m & 1.13$\pm$0.02 & 1.14$\pm$0.02 & 1.17$\pm$0.04 \\
\hline
Multiscale grid 1 & 0.43m & 1.12$\pm$0.02 & 1.17$\pm$0.02 & 1.19$\pm$0.03 \\
\hline
Multiscale grid 2 & 0.30m & 1.25$\pm$0.02 & 1.25$\pm$0.02 & 1.23$\pm$0.03 \\
\hline
\end{tabular}
\caption{Estimates of the energy decay exponent $n$ obtained using three different methods. Note that the uncertainty in $n$ indicated above for a given method does not represent the statistical uncertainty in $n$, but rather the range of values obtained using a given procedure for a randomly chosen set of data.}
\end{center}
\label{tab1}
\end{table}
%

\subsection{\label{Local}Local exponent method}

Measurements of $\left< u_x^2 \right>$ were made at intervals of $\Delta x = 200mm$ or $\Delta x \approx 8.5 \ell_0$ with a positioning accuracy better than $\epsilon_x = 1mm$ and so local exponents can be estimated using two-point differencing applied to Eq. (\ref{Eq4-7}):
\begin{equation}
n = - \frac{{\rm ln} \left[ \frac{\left< u_x^2\right> \left( x + \Delta x - x_0 \right)} {\left< u_x^2\right> \left( x - \Delta x - x_0 \right) } \right]}{{\rm ln} \left[ \frac{ \left( x + \Delta x - x_0 \right)} { \left( x - \Delta x - x_0 \right) } \right]}  \; .
\label{Eq4-8}
\end{equation}

\begin{figure}
\begin{center}
  \hspace{-3mm}
  \includegraphics[width=1.0\textwidth]{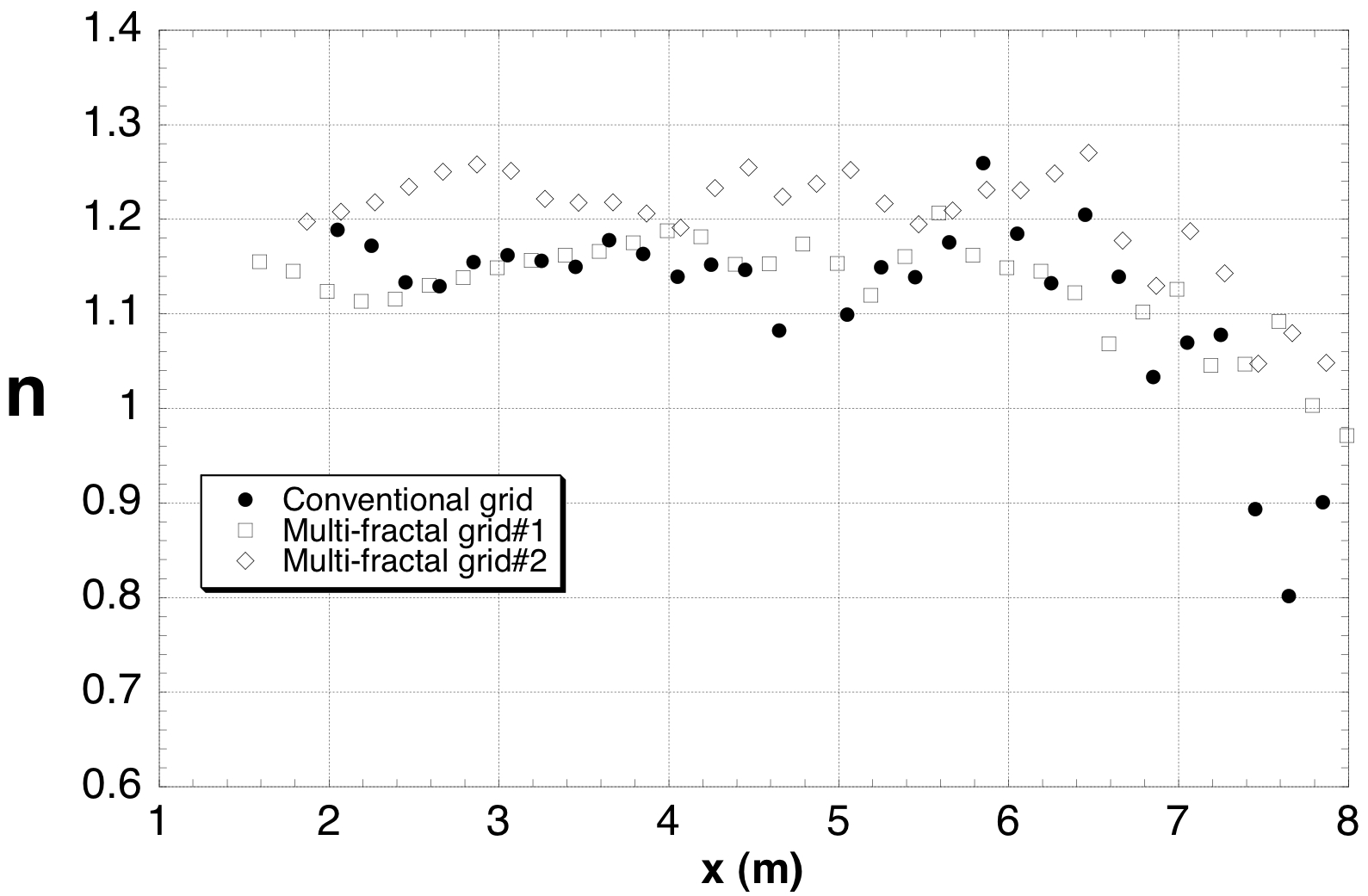}
  \caption{Local exponents, $n$, obtained for $\left< u_x^2 \right>$ using Eq. (\ref{Eq4-8}).}
 \label{fig:Local}
 \end{center}
\end{figure}

This procedure has the advantage that the unknown coefficient $a$ is eliminated, but the disadvantage that it is sensitive to noise. Using the values of $x_0$ obtained from the regression method, the estimates of $n$ obtained using Eq. (\ref{Eq4-8}), as applied to $\left< u_x^2 \right>$, are shown in Figure \ref{fig:Local}. It is clear that these estimates of $n$ are sensitive to noise for $x$ greater than, say, $6.5m$. Never-the-less, reasonably constant values are found for the range $2m < x < 6.5m$, and averaging over this range provides a single estimate of the exponent $n$ for each grid. Of course, this estimate varies slightly depending on whether data for $\left< u_x^2 \right>$ or $\left< q^2 \right>$ is analyzed, and on the precise range of $x$ used to form the average. The range of exponents thus obtained for each grid is also shown in Table 1. It is reassuring that these values are close to those found by the regression method.
%

\subsection{\label{MaxDecay}Maximum decay range method}

In this method $x_0$ is chosen to give the longest power-law range in $x$ and then $n$ is determined for that value of $x_0$. The procedure works as follows. First $x_{max}$ is fixed at the value given by the regression method (at around 8m). Next, a series of different values of $x_0$ are chosen, and for each $x_0$ the best fit value of $n$ is obtained for a range of values of $x_{min}$. The value of $x_0$ which gives the widest range of constant $n$ between $x_{min}$ and $x_{max}$ is then deemed to be the correct one. The exponent $n$ then follows.

\begin{figure}
\begin{center}
  \hspace{-3mm}
  \includegraphics[width=1.0\textwidth]{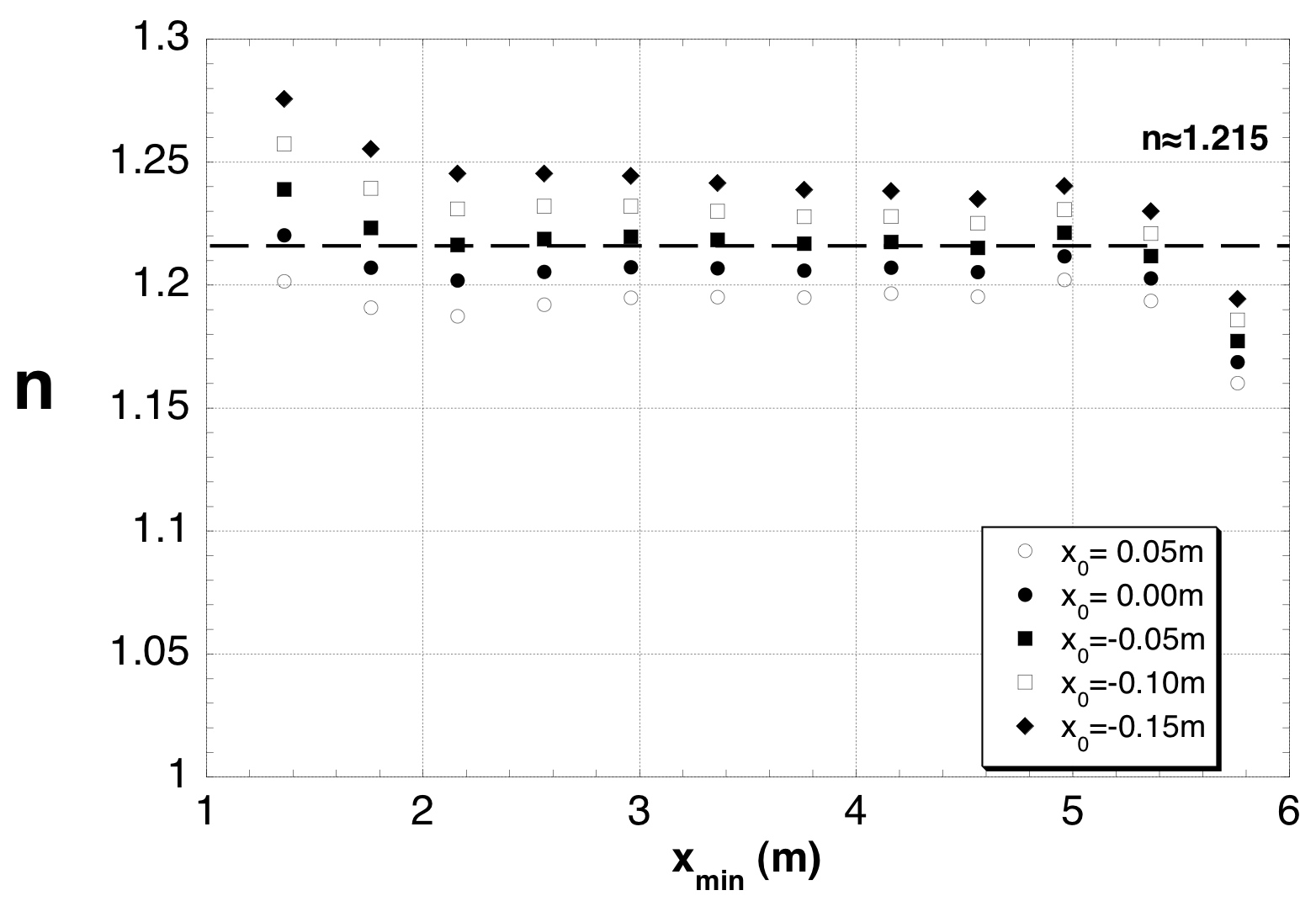}
  \caption{Decay exponents, $n$, for the grid $msg2$ as a function of $x_{min}$ for different values of the virtual origin $x_0$.}
 \label{fig:Bob}
 \end{center}
\end{figure}

The application of this method to the conventional grid is shown in \cite{Krogstad} and demonstrated here by applying it to the $\left< q^2 \right>$ data from $msg2$ in Figure \ref{fig:Bob}. As with the other methods, slightly different values of $n$ are obtained for a given grid, depending on whether $\left< u_x^2 \right>$ or $\left< q^2 \right>$ are analyzed. The values obtained are given in ranges indicated in Table 1, where it can be seen that all three procedures generate similar exponents.

There are three striking features of Table 1. First, all three grids yield decay exponents very close to the classical Saffman value of $n = 6/5$. Second, the conventional and multi-scale grids produce almost identical results (to within experimental uncertainty). Third, these decay exponents are a long way from the $n$ in the range 1.7 to  2.0 quoted in \cite{Hurst} for their four fractal cross grids. It is likely, therefore, that all three of our grids produce Saffman turbulence.	

\begin{figure}
\begin{center}
  \hspace{-3mm}
  \includegraphics[width=1.0\textwidth]{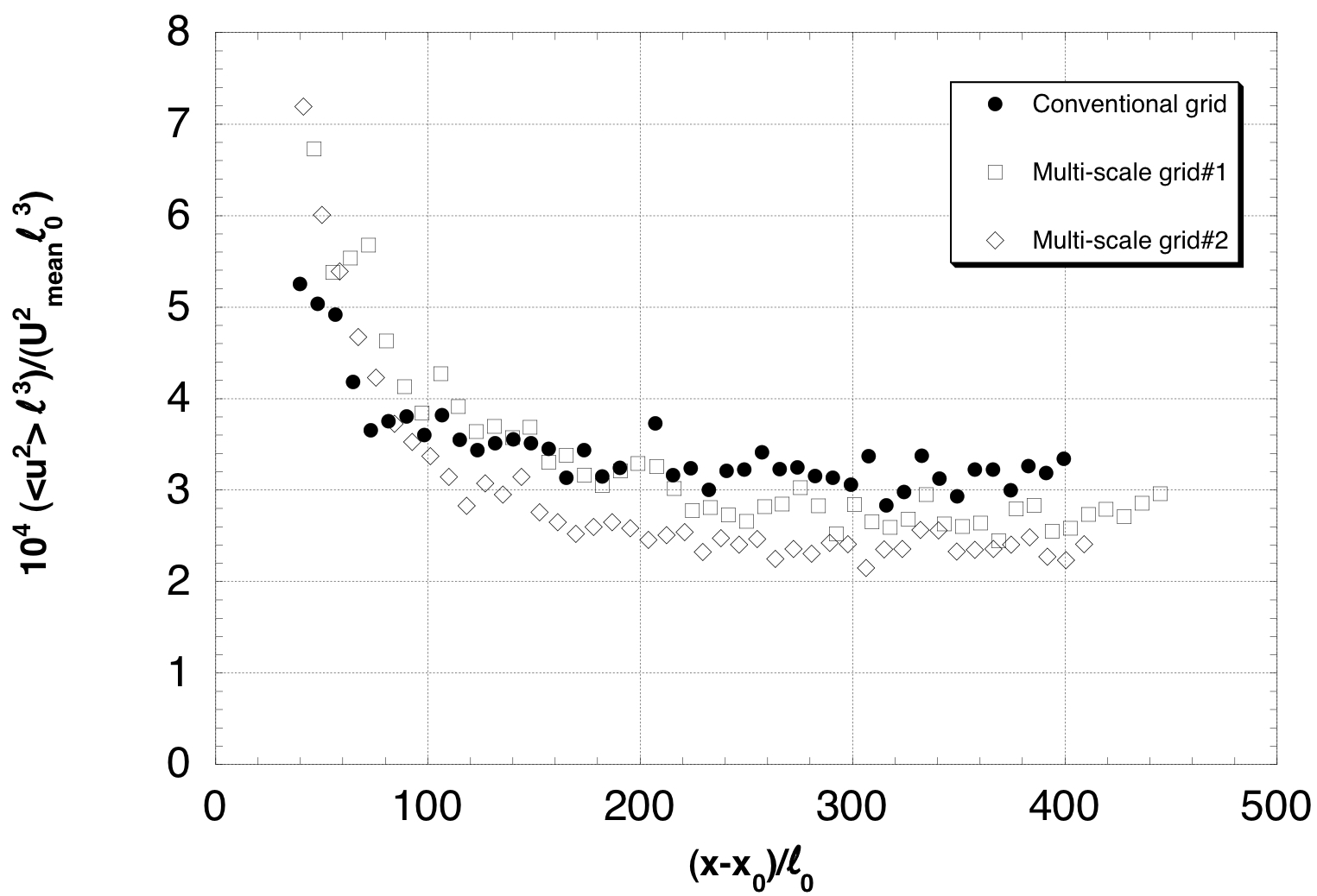}
  \caption{$\left< u_x^2 \right> \ell^3 / U^2 \ell_0^3$ versus $(x - x_0) / \ell_0$ for all three grids.}
 \label{fig:Saffman}
 \end{center}
\end{figure}

We can check the hypothesis that we have Saffman turbulence if we pre-empt our discussion of length scales in $\S$ \ref{Development}. Figure \ref{fig:Saffman} shows $\left< u_x^2 \right> \ell^3 / U^2 \ell_0^3$ versus $(x - x_0) / \ell_0$ for all three grids. Recall that, in Saffman turbulence, self-similarity of the large scales requires $u^2 \ell^3$ = constant, as distinct from, say, Batchelor turbulence, in which $u^2 \ell^5$ = constant. It is clear from Figure \ref{fig:Saffman} that there is an initial transient, which is more or less restricted to $(x - x_0) / \ell_0 < 100$ ($x < 2.7m$), after which $u^2 \ell^3$ is indeed more or less constant in all three cases. The scatter in this data is largely a consequence of the difficulty associated with estimating $\ell$, as will be discussed in $\S$ \ref{Development}. 

\begin{figure}
\begin{center}
  \hspace{-3mm}
  \includegraphics[width=1.0\textwidth]{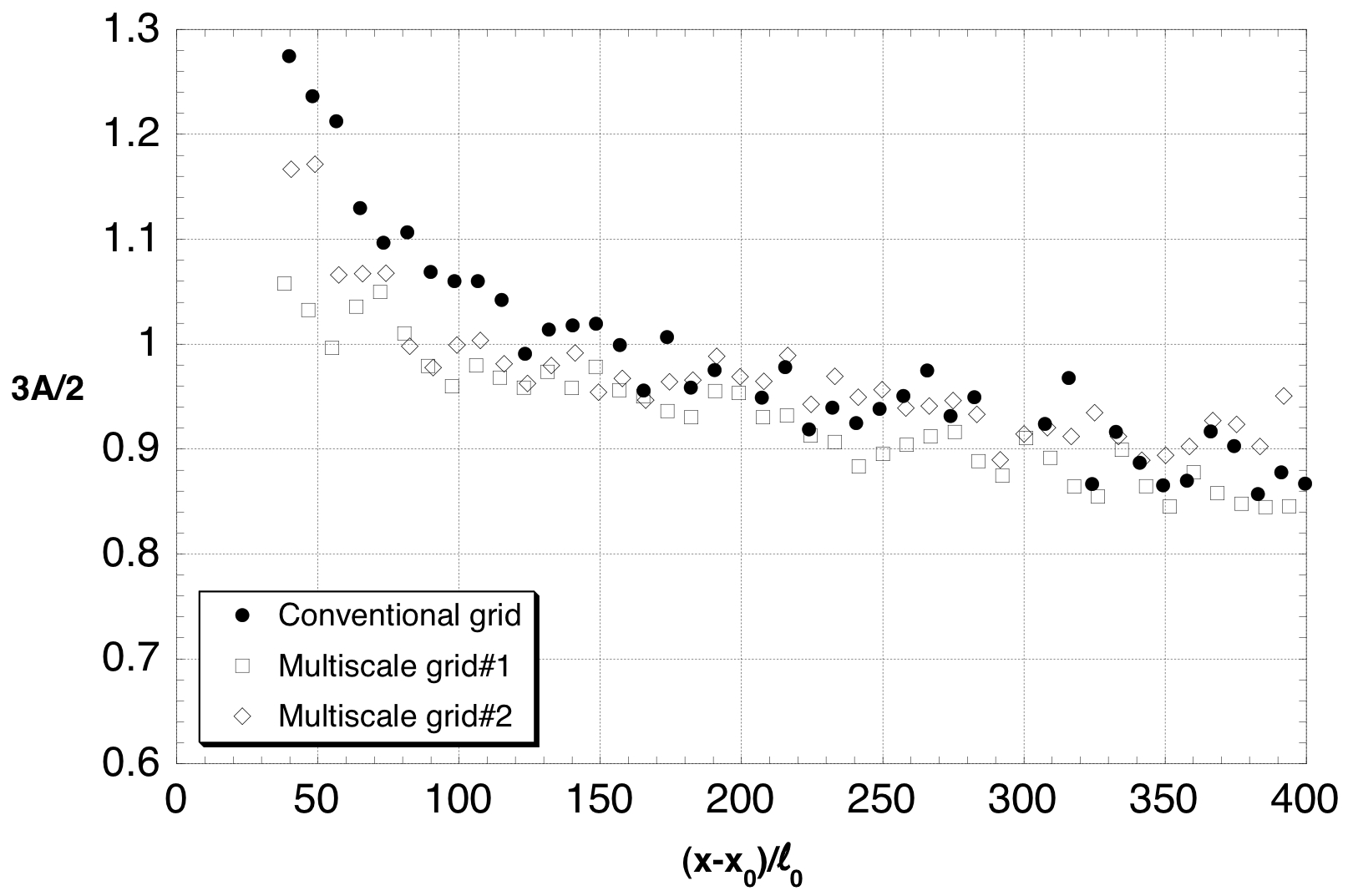}
  \caption{Streamwise development of $A$ obtained from Eq. (\ref{Eq4-9}).}
 \label{fig:A}
 \end{center}
\end{figure}

Finally we consider the dimensionless dissipation coefficient $A$ in Eq. (\ref{Eq4-5}), which is normally taken to be constant during the decay of isotropic turbulence. Assuming isotropy at the small scales, the viscous dissipation rate, $\epsilon$, can be written as
\begin{equation}
\epsilon = \frac{3}{2} \frac{A u^3}{\ell} = 15 \nu \left< \left( \frac{\partial u_x}{\partial x}Ê\right)^2 \right>  \; ,
\label{Eq4-9}
\end{equation}
which allows us to estimate $A$ from measurements of $\left< \left( \partial u_x / \partial xÊ\right)^2 \right>$, $u$ and $\ell$. The corresponding values of $A$ are plotted in Figure \ref{fig:A} for all three grids as a function of $(x - x_0) / \ell_0$. Again, there is some scatter, largely due to the difficulty in estimating $\ell$. Never-the-less, once the turbulence is fully developed, say for $(x - x_0) / \ell_0 > 100$, there is a slow but steady decline in $A$ which is consistent across all three grids. As noted earlier, this slow variation in $A$ means that, even if we have Saffman turbulence, with $u^2 \ell^3$ = constant, we need not recover $n = 6 / 5$, as this value of $n$ relies on $A$ being strictly constant.
%

\section{\label{Development}The Streamwise Development of Length Scales}

\begin{figure}
\begin{center}
  \hspace{-3mm}
  \includegraphics[width=1.0\textwidth]{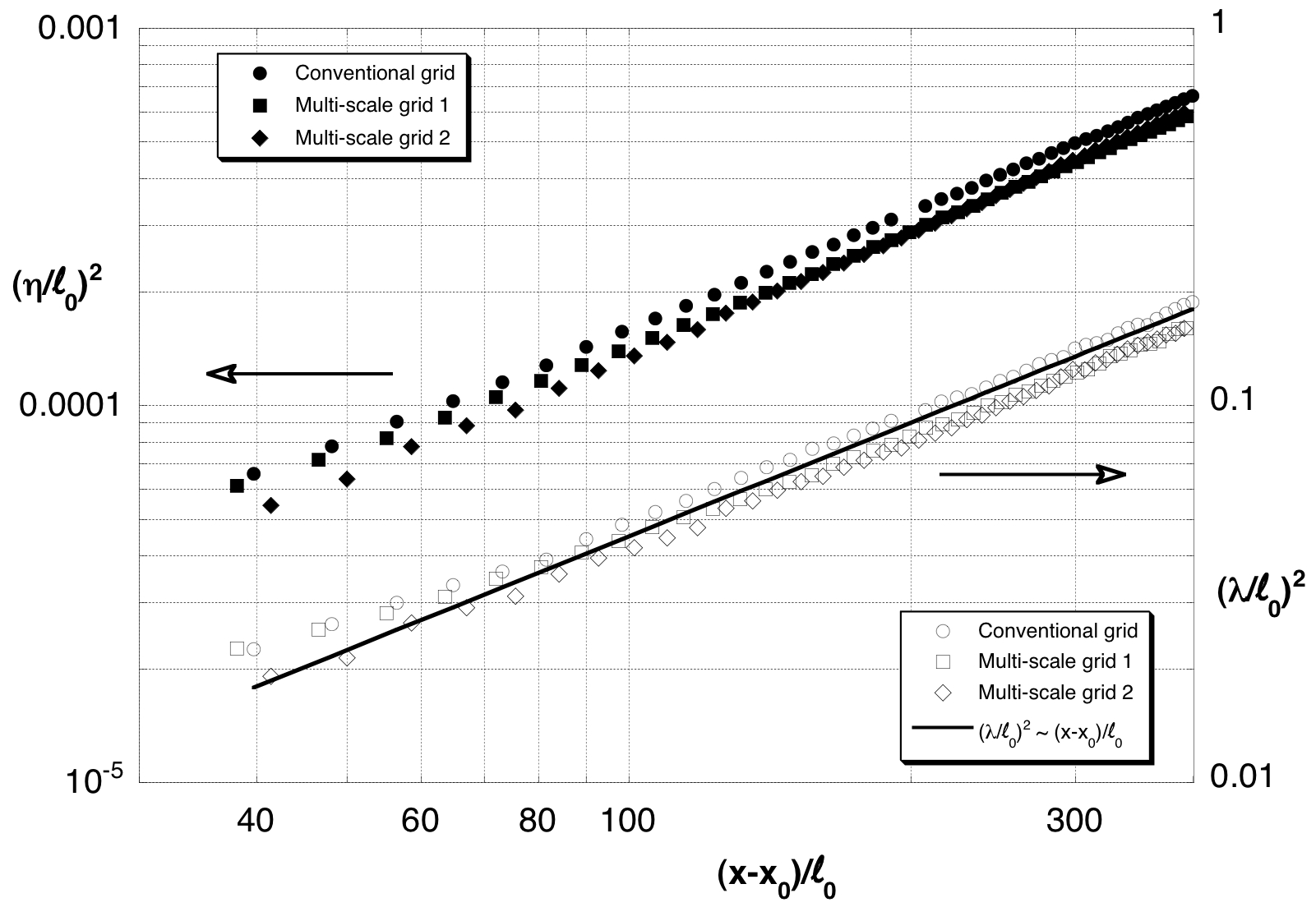}
  \caption{Streamwise distributions of $\eta^2$ (filled symbols) and $\lambda^2$ (open symbols). The straight line is $\lambda^2 \sim (x - x_0)$.}
 \label{fig:Smallscales}
 \end{center}
\end{figure}

We now turn to the length scales $\eta$, $\lambda$ and $\ell$. The Kolmogorov microscale, defined as $\eta = \left( \nu^3 / \epsilon \right)^{1/4}$, can be determined from the isotropic estimate of dissipation in Eq. (\ref{Eq4-9}) or from the decay rate of $\left< q^2 \right>$, both estimates giving virtually identical estimates for $\eta$ in this type of flow, as demonstrated by \cite{Krogstad}. The streamwise development of the Kolmogorov and Taylor microscales, $\eta^2$  and $\lambda^2$, are shown in Figure \ref{fig:Smallscales}.

Noting that combining Eq. (\ref{Eq4-5}) and (\ref{Eq4-6}) yields 
\begin{equation}
\lambda^2 = \frac{10 \nu \ell}{A u} =  \frac{10 \nu (x - x_0 )}{n U}   \; ,
\label{Eq5-1}
\end{equation}
we see that $\lambda^2$ should scale as $\lambda^2 \sim (x - x_0 )$, which is indeed verified in Figure \ref{fig:Smallscales}. This is, in effect, confirmation of a power-law form of energy decay.

\begin{figure}
\begin{center}
  \hspace{-3mm}
  \includegraphics[width=1.0\textwidth]{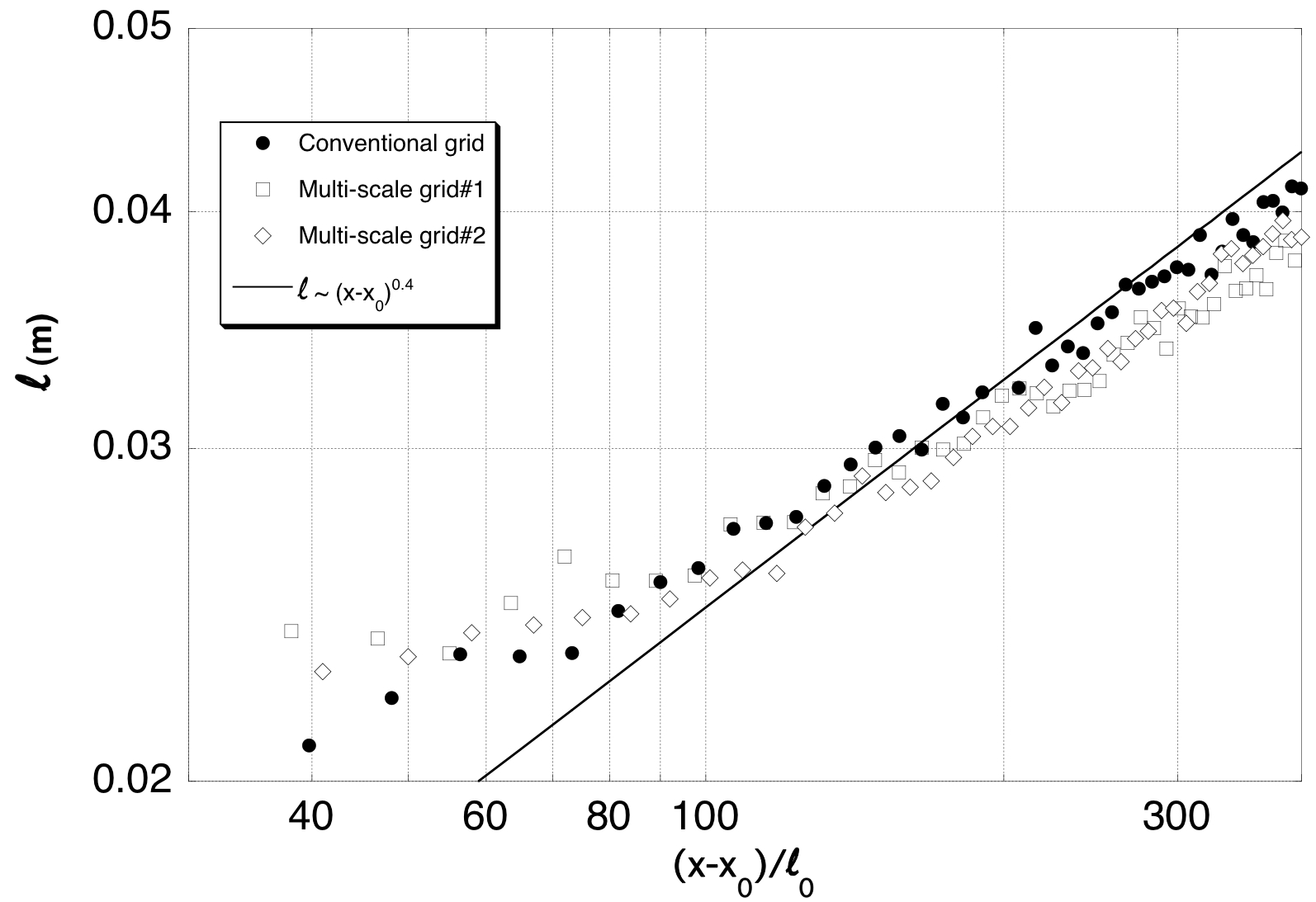}
  \caption{The integral scale, $\ell$, as a function of $(x - x_0 ) / \ell_0$. The solid line is $\ell \sim (x - x_0 )^{0.4}$.}
 \label{fig:Largescales}
 \end{center}
\end{figure}

As noted in $\S$ \ref{Exp}, the integral scale, $\ell$, is defined in the usual way as
\begin{equation}
\ell = \int_0^{\infty} \frac{\left< u_x (x) u_x (x+r) \right>}{\left< u_x^2 \right>} dr=   \int_0^{\infty} f(r) dr   \; ,
\label{Eq5-2}
\end{equation}
where $f(r)$ is the usual longitudinal correlation function and, in practice, $\left< u_x (x) u_x (x+r) \right>$ is evaluated using Taylor's hypothesis. Ideally $f(r)$ should decay monotonically to zero for large $r$, but in experiments $f(r)$ almost always exhibits a weak oscillatory tail which persists for many multiples of $\ell$. This makes it difficult to evaluate $\int_0^{\infty} f(r) dr$ accurately since it will depend on where the integral is terminated, and so the integral in Eq. (\ref{Eq5-2}) was only taken up to the first zero crossing, which introduces a small systematic error. As a result, there is some uncertainty in the calculated values of $\ell$. Our estimates of $\ell$ for all three grids are shown in Figure \ref{fig:Largescales}.

Evidently, $\ell$ exhibits considerably more scatter than $\eta$ or $\lambda$, but it is clear that the development of $\ell$ is similar for all three grids and follows a power law. Moreover,  power-law energy decay with constant $A$ demands $\ell \sim u t$, and so Saffman turbulence predicts $\ell \sim (x - x_0 )^{0.4}$, provided we ignore the slow decline in $A$. This is shown in Figure \ref{fig:Largescales} for comparison and the data for all cases follow such a trend for $(x - x_0 ) / \ell > 100$.

\section{\label{Spectra}Spectral Development of the Turbulence}

We have seen that, as far as the behavior of the integral scales $u$ and $\ell$ are concerned, the conventional and multi-scale grids behave in almost identical ways. However, it might be argued that, because energy is injected across the scales with a multi-scale grid, the evolution of the energy spectrum may exhibit non-classical features. So we conclude by comparing the spectra of the multi-scale and conventional grids. We shall focus on the usual one-dimensional spectrum, $F_{11}(k)$, defined by the transform pair
\begin{equation}
F_{11}(k) = \frac{2}{\pi}\int_0^{\infty} \left< u_x (x) u_x (x+r) \right> {\rm cos} (kr) dr  \; ,
\label{Eq6-1}
\end{equation}
\begin{equation}
\left< u_x (x) u_x (x+r) \right> = \int_0^{\infty} F_{11}(k) {\rm cos} (kr) dk   \; .
\label{Eq6-2}
\end{equation}

\begin{figure}
\begin{center}
  \hspace{-3mm}
\subfigure{
  \includegraphics[width=.47\textwidth]{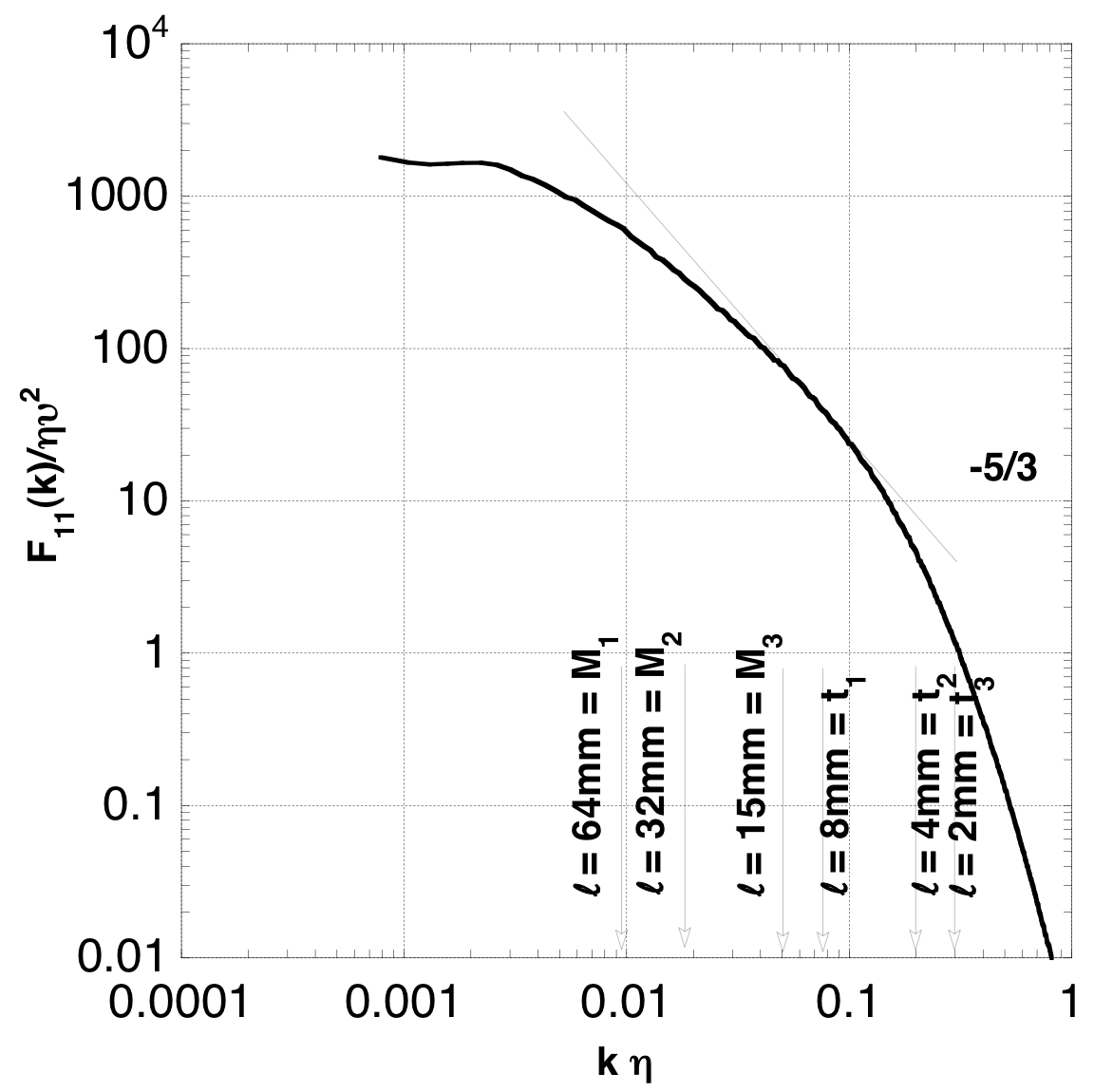}
  \label{fig:Fig13a}}
\hspace{2mm}
\subfigure{
  \includegraphics[width=.47\textwidth]{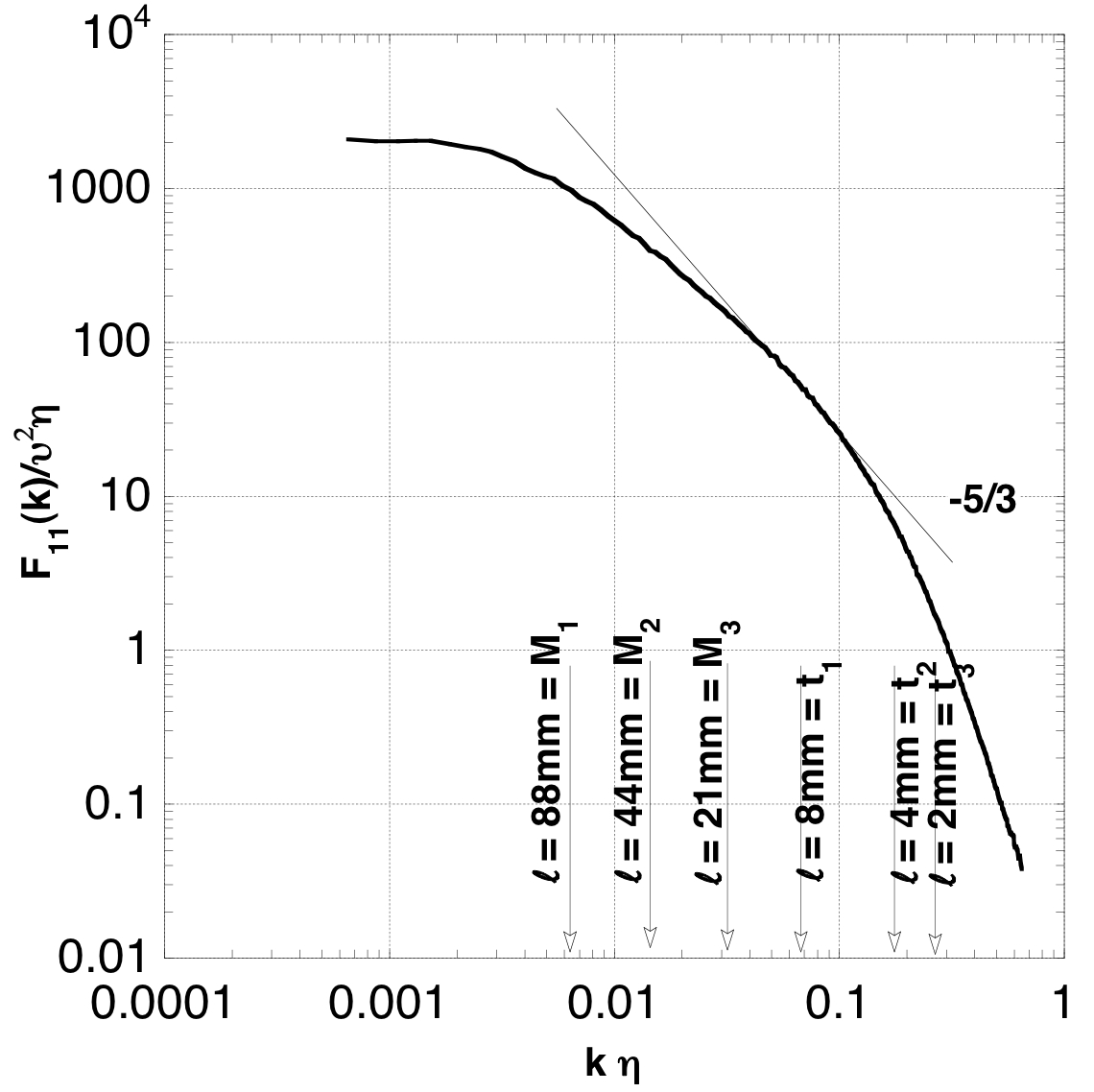}
  \label{fig:Fig13b}}
  \vspace{-2mm}
  \hspace{-4mm}
(a)
  \hspace{61mm}
(b)
  \caption{$F_{11}(k)$, normalized by the Kolmogorov microscales $\eta$ and $\upsilon$, for (a) $msg1$ and (b) $msg2$, measured near the grid at $x \approx 1.3m$. }
 \label{fig:Spectra-x0}
 \end{center}
\end{figure}

Let us first consider the issue of the energy injection scales for the multi-scale grids. Figure \ref{fig:Spectra-x0} shows $F_{11}(k)$ for these two grids, measured just downstream of the mesh (at $x \approx 1.3m$), and normalized by the Kolmogorov microscales, $\eta$ and $\upsilon$. The mesh sizes $M_i$ and bar widths $t_i$, are marked on the horizontal axis using the rule of thumb that $k = \pi / \ell$. It is clear that in both cases the range of geometric scales spans most of the turbulent spectrum.

\begin{figure}
\begin{center}
  \hspace{-3mm}
\subfigure{
  \includegraphics[width=.46\textwidth]{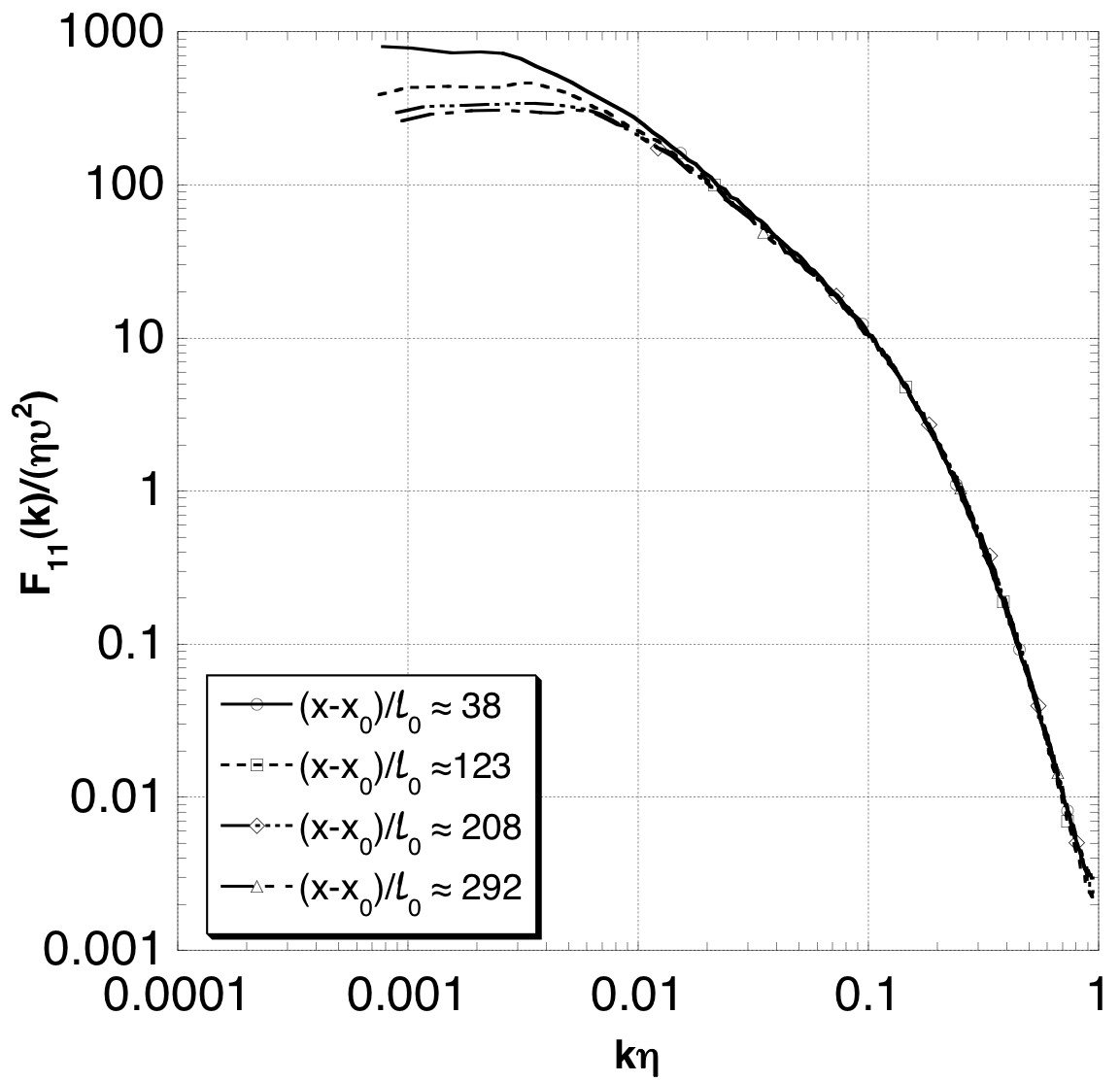}
  \label{fig:Fig14a}}
\hspace{2mm}
\subfigure{
  \includegraphics[width=.49\textwidth]{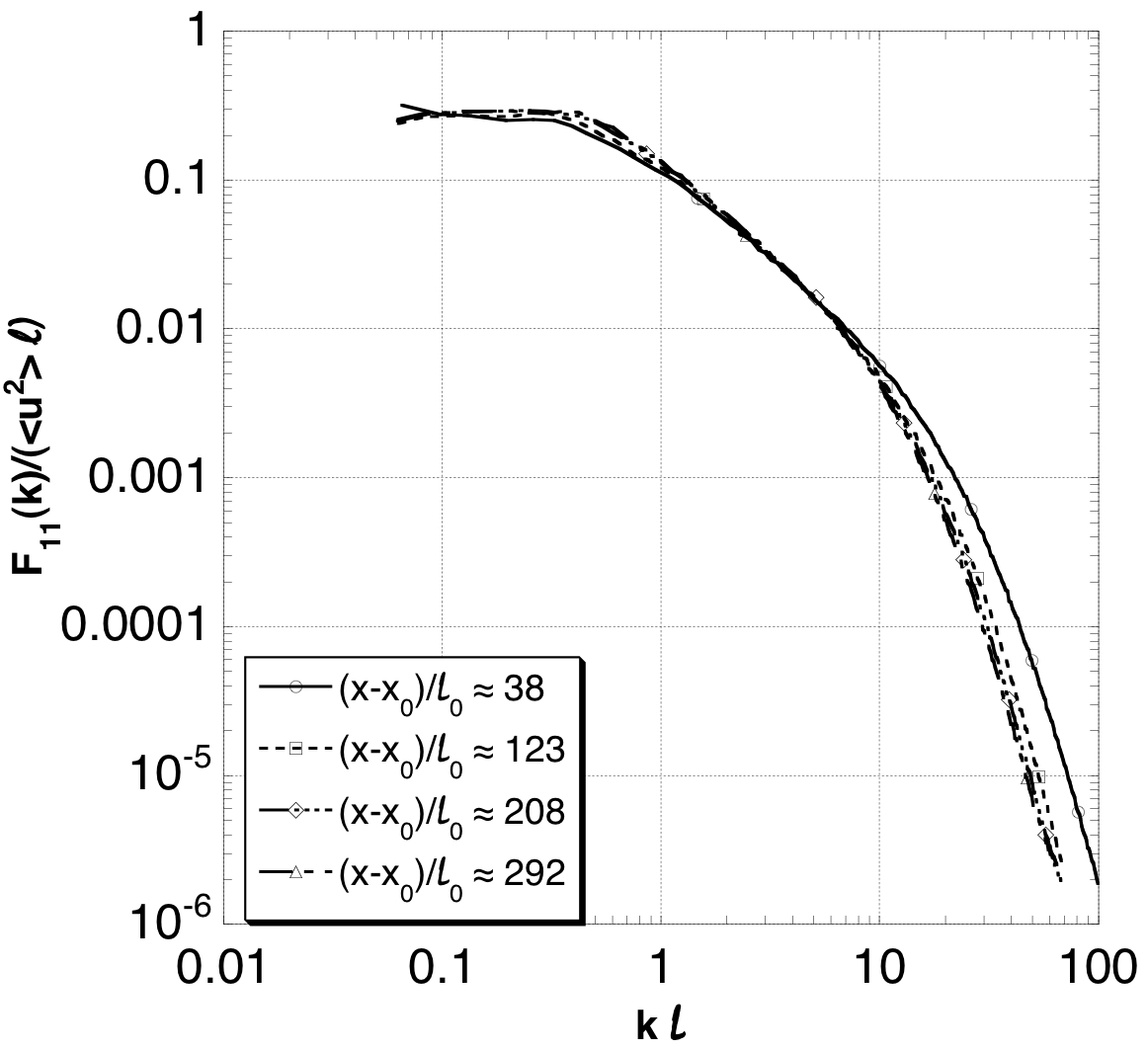}
  \label{fig:Fig14b}}
  \vspace{-2mm}
  \hspace{-4mm}
(a)
  \hspace{61mm}
(b)
  \caption{$F_{11}(k)$ for grid $msg1$ measured at four streamwise locations and normalized by (a) $\eta$ and $\upsilon$, and (b) $\ell$ and $u$. }
 \label{fig:Spectra-grid1}
 \end{center}
\end{figure}

\begin{figure}
\begin{center}
  \hspace{-3mm}
\subfigure{
  \includegraphics[width=.46\textwidth]{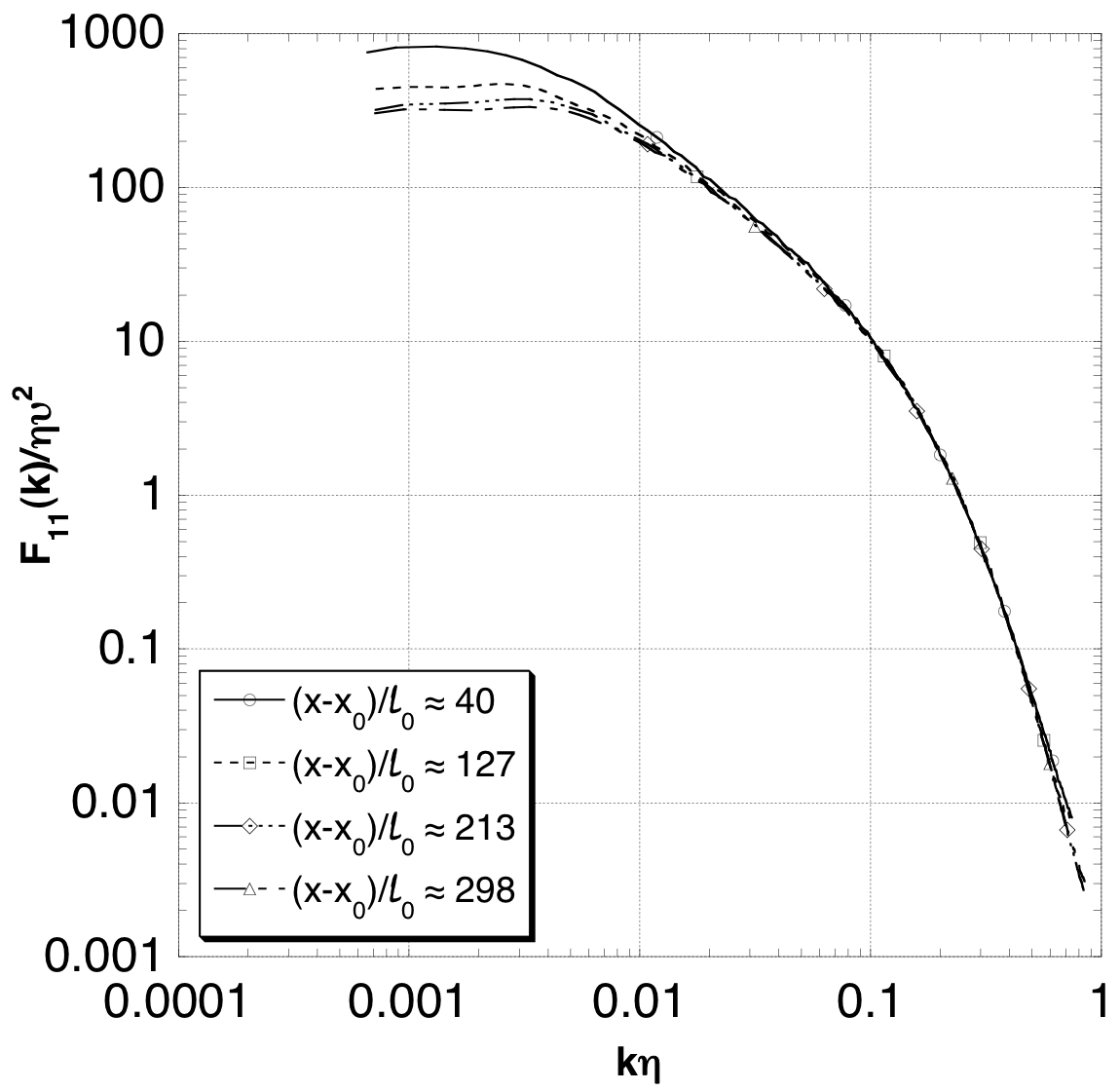}
  \label{fig:Fig15a}}
\hspace{2mm}
\subfigure{
  \includegraphics[width=.49\textwidth]{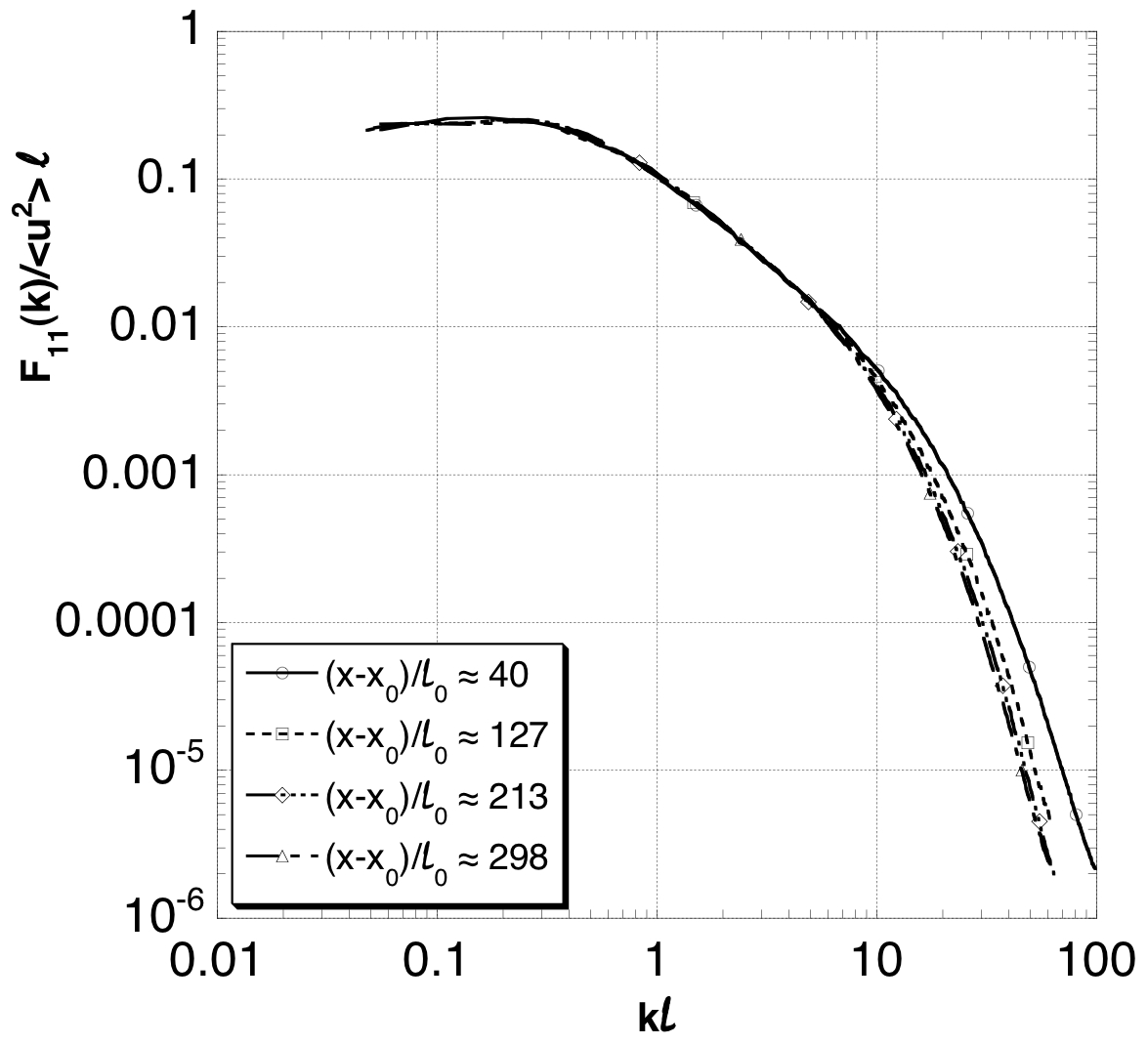}
  \label{fig:Fig15b}}
  \vspace{-2mm}
  \hspace{-4mm}
(a)
  \hspace{61mm}
(b)
  \caption{$F_{11}(k)$ for grid $msg2$ measured at four streamwise locations and normalized by (a) $\eta$ and $\upsilon$, and (b) $\ell$ and $u$. }
 \label{fig:Spectra-grid2}
 \end{center}
\end{figure}

Next there is the issue of whether or not the turbulence produced by the multi-scale grids behaves in the classical way, with the spectrum collapsing at high $k$ when normalized by the microscales $\eta$ and $\upsilon$, and at low $k$ when normalized by the integral scales $\ell$ and $u$. Figure \ref{fig:Spectra-grid1} shows $F_{11}(k)$ for grid $msg1$ measured at four different streamwise locations and normalized by (a)   and $\eta$ and $\upsilon$ and (b) $\ell$ and $u$. In case (a) there is excellent collapse of the spectra at high $k$ and no collapse at low $k$, while in (b) there is reasonable collapse at low $k$ and no collapse at high $k$. The same plots are given for $msg2$ in Figure \ref{fig:Spectra-grid2} and again there is excellent collapse at high $k$ when normalized by $\eta$ and $\upsilon$, and at low $k$ when scaled by $\ell$ and $u$ also in this case. Once again, we see that the turbulence produced by the multi-scale grids behaves in a classical manner.

\begin{figure}
\begin{center}
  \hspace{-3mm}
\subfigure{
  \includegraphics[width=.47\textwidth]{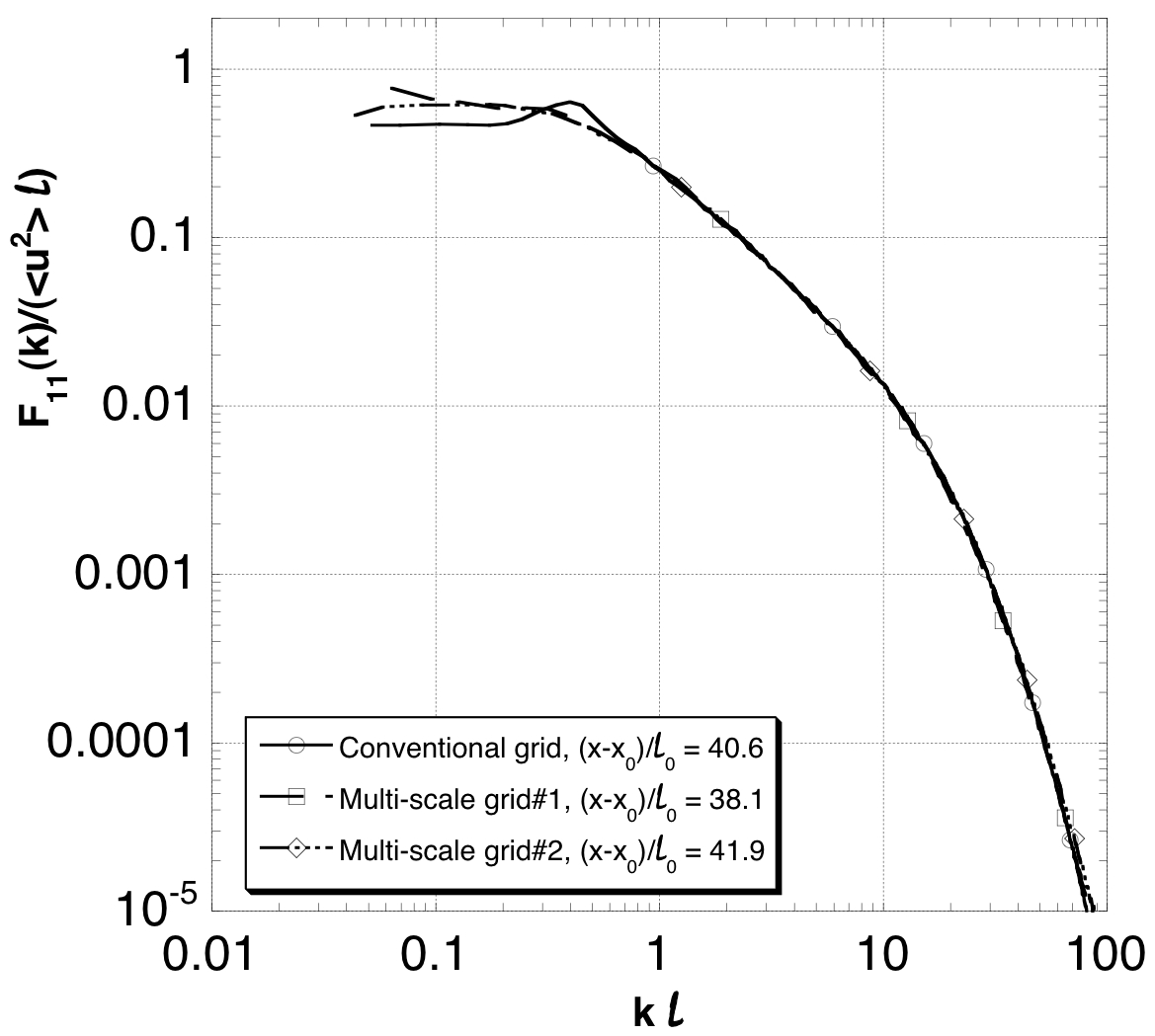}
  \label{fig:Fig16a}}
\hspace{4mm}
\subfigure{
  \includegraphics[width=.46\textwidth]{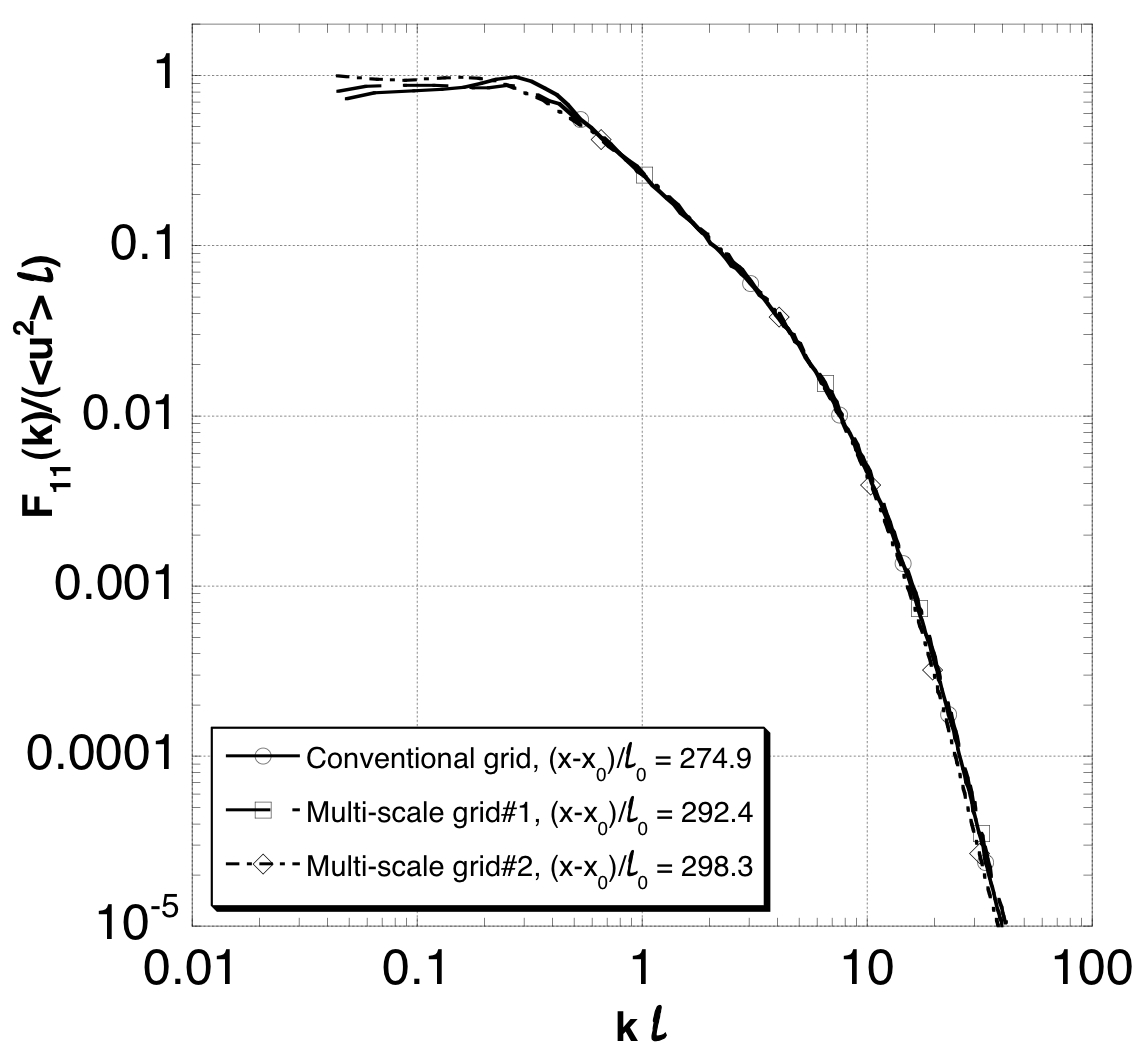}
  \label{fig:Fig16b}}
  \vspace{-2mm}
  \hspace{-4mm}
(a)
  \hspace{61mm}
(b)
  \caption{$F_{11}(k)$ for all three grids, normalized by the integral scales $\ell$ and $u$ at (a) $(x - x_0 ) / \ell_0 \approx 40$ and (b) $(x - x_0 ) / \ell_0 \approx 290$. }
 \label{fig:Spectra-all}
 \end{center}
\end{figure}

\begin{figure}
\begin{center}
  \hspace{-3mm}
\subfigure{
  \includegraphics[width=.47\textwidth]{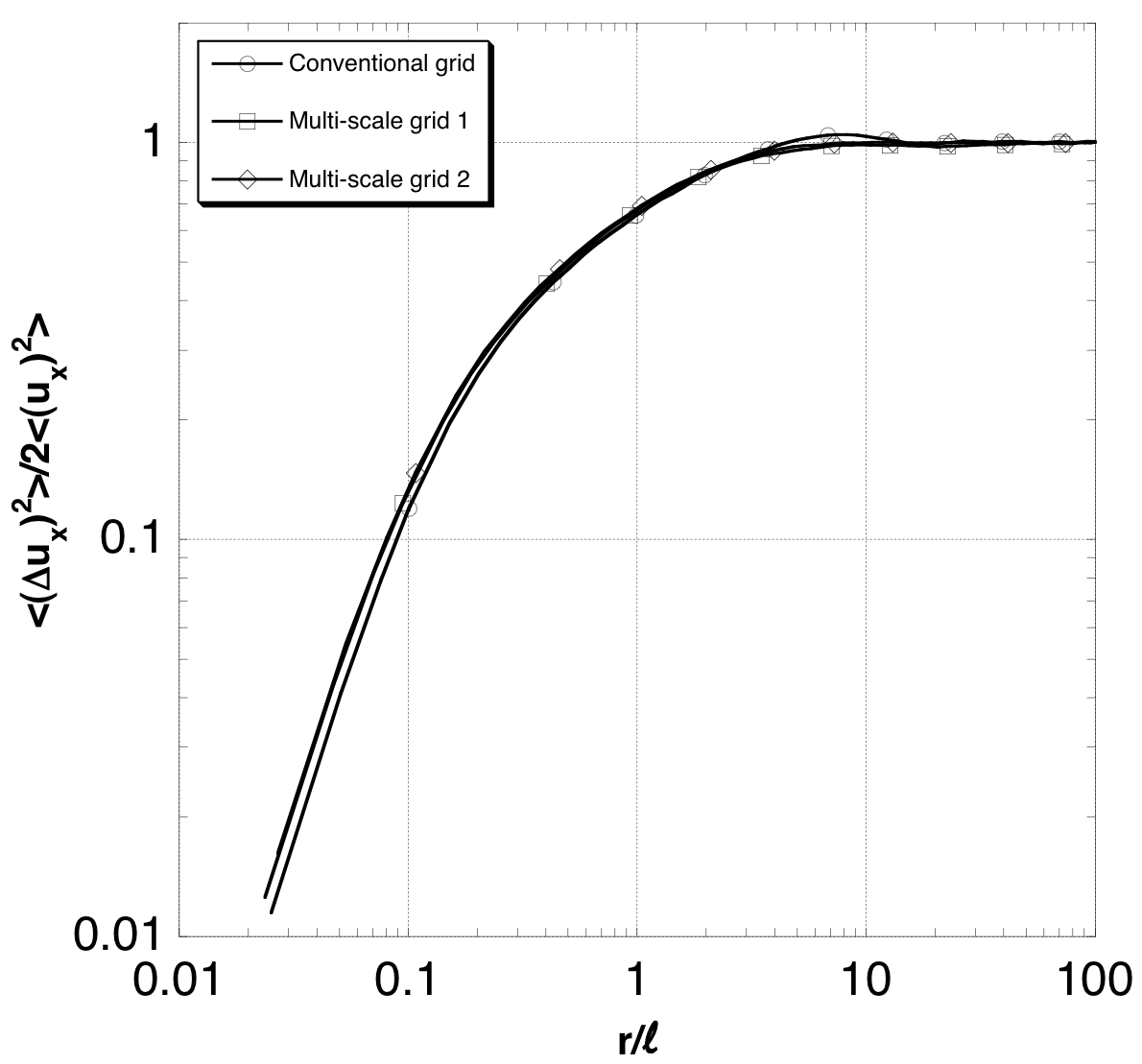}
  \label{fig:Fig17a}}
\hspace{4mm}
\subfigure{
  \includegraphics[width=.46\textwidth]{Fig18a}
  \label{fig:Fig17b}}
  \vspace{-2mm}
  \hspace{-4mm}
(a)
  \hspace{61mm}
(b)
  \caption{$\left< \left( \Delta u_x \right)^2 \right> / 2\left< u_x^2 \right>$ plotted against $r/\ell$ for all three grids at (a) $(x - x_0 ) / \ell_0 \approx 40$ and (b) $(x - x_0 ) / \ell_0 \approx 290$.}
 \label{fig:Strufun}
 \end{center}
\end{figure}

Finally we compare spectra at the same location generated by the three different grids. Figure \ref{fig:Spectra-all} shows $F_{11}(k)$ for all three grids, normalized by the integral scales $\ell$ and $u$, at (a) $\ell \sim (x - x_0 ) / \ell_0 \approx 40$ and (b) $\ell \sim (x - x_0 ) / \ell_0 \approx 290$. There is excellent collapse at both streamwise locations, confirming that the nature of the turbulence coming off the conventional and multi-scale grids is essentially the same. The same information may be examined in real space using the longitudinal structure function 
\begin{equation}
\left< \left( \Delta u_x \right)^2 \right> = \left< \left(u_x (x+r) - u_x (x) \right)^2 \right> = 2 \left< u_x^2 \right> - 2 \left< u_x (x) u_x (x+r)  \right>   \; ,
\label{Eq6-3}
\end{equation}
which, crudely speaking, acts as a cumulative index of all energy held in structures smaller than scale $r$. Figure \ref{fig:Strufun} shows $\left< \left( \Delta u_x \right)^2 \right> / 2\left< u_x^2 \right>$ plotted against $r/\ell$ at the same two streamwise locations as Figure \ref{fig:Spectra-all}. Again it can be seen that there is good collapse at both locations. 

\section{Conclusions}

Our primary findings are two-fold. First, it seems that Saffman's decay law is reasonably robust, since the energy decay exponents for all three grids are close to Saffman's classical prediction of $n$ = 6/5. Second, the multi-scale grids used here produce almost identical results to the equivalent classical grid. In particular, all three flows exhibit remarkably similar streamwise distributions of $Re_\lambda$ (Figure \ref{fig:Re_lambda}), flatness and skewness (figure 5), and dimensionless decay coefficient $A$ (Figure \ref{fig:A}). It is also worth noting that and the spectra for the multiscale grids exhibit classical Kolmogorov scaling, with $E(k)$ collapsing on $\ell$ and $u$ at low $k$, and on $\eta$ and $\upsilon$ at high $k$.

Our findings contradict those of some previous studies which report unusual behavior behind similar multi-scale grids, in particular, a very high energy decay exponent of around $n \sim$ 2.0  and unusually high values of $Re_\lambda$. A decay exponent of $n \sim$ 2.0 is particularly worrying as the theoretical maximum for $n$ (assuming the dimensionless decay coefficient, $A$, is constant) is $n$ = 10/7. However, these earlier measurements were taken much closer to the grid where the flow exhibits initial grid-dependent inhomogeneities; inhomogeneities which, according to the present data, disappear further downstream.

\appendix
\section{\label{appendix}Classical decay exponents}

The classical decay exponents of isotropic turbulence, including those of \cite{Kolmogorov} and \cite{Saffman}, are obtained as follows. (See \citeauthor{Ossai}, \citeyear{Ossai}, \citeauthor{Ishida}, \citeyear{Ishida}, \citeauthor{davidson04}, \citeyear{davidson04} and \citeauthor{davidson09}, \citeyear{davidson09}, for more details.) We start with the result that, in fully-developed turbulence, $E\left( k \rightarrow 0 \right) = c_m k^m$, where $c_m \sim u^2 \ell^{m+1}$ and $c_m$ = constant for $m \leq 4$. Self-similarity of the large scales then demands $u^2 \ell^{m+1}$ = constant for $m \leq 4$. When combined with the zeroth law, Eq. \ref{Eq4-5}, this yields
\begin{equation}
 \frac{u^2}{u^2_0} = \left[ 1 + \frac{A}{n} \frac{u_0 t}{\ell_0}Ê\right]^{-n} \;\; ,\;\; \;\; n = 2 - \frac{4}{m+3} \;\; ,
\label{EqA-1}
\end{equation}
where the subscripts 0 indicate values at $t = 0$. The smallest value of $m$ is $m = 2$, since lower values require that the spectral tensor diverges as $k \rightarrow 0$. On the other hand, the largest value is $m = 4$, as higher values spontaneous convert to $m = 4$ during an initial transient. Thus the classical range of decay exponents is, $6/5 \leq n \leq 10/7$.


\bibliographystyle{jfm}

\end{document}